\documentclass[final,5p,times,twocolumn]{elsarticle}

\usepackage{amssymb}
\usepackage{amsmath}
\journal{Chaos Solitons $\&$ Fractals}

\begin{document}

\begin{frontmatter}



\title{Modulation Instability-Induced Multimode Squeezing in Quadratic Frequency Combs} %% Article title


\affiliation[add1]{organization={State Key Laboratory of Photonics and Communications, School of Information Science and Electronic Engineering $\&$ School of Integrated Circuits}, 
            addressline={Shanghai Jiao Tong University}, 
            city={Shanghai},
            postcode={200240}, 
            country={China}}
            
\affiliation[add2]{organization={State Key Laboratory of Precision Spectroscopy}, 
            addressline={East China Normal University}, 
            city={Shanghai},
            postcode={200062}, 
            country={China}}
            
\affiliation[add3]{organization={Microwave Photonics Technology Key Laboratory of Sichuan Province}, 
            city={Chengdu},
            state={Sichuan Province}, 
            country={China}}
            
\affiliation[add4]{organization={School of Physics and Optoelectronic Engineering, Institute of Laser Engineering}, 
            addressline={Beijing University of Technology}, 
            city={Beijing},
            postcode={100124}, 
            country={China}}
            
\affiliation[add5]{organization={College of Electronic Science, National key laboratory of satellite navigation technology}, 
            addressline={National University of Defense Technology}, 
            city={Changsha},
            postcode={410073}, 
            country={China}}
            
\affiliation[add6]{organization={National Key Laboratory of Metrology and Calibration}, 
            addressline={Beijing Changcheng Institute of Metrology $\&$ Measurement}, 
            city={Beijing},
            postcode={100095}, 
            country={China}}

% 作者列表与单位关联
% 作者列表（添加共同一作标记t1）
\author[add1]{Haodong Xu\fnmark[t]}  % 第一位作者添加t1标记
\author[add1]{Nianqin Li\fnmark[t]}  % 第二位作者添加t1标记
\author[add1]{Zijun Shu}
\author[add1]{Yang Shen}
\author[add1]{Bo Ji}
\author[add3]{Aiping Xie}
\author[add4]{Feng Yang}
\author[add4]{Dengcai Yang}
\author[add5]{Jing Peng}
\author[add5]{Hang Gong}
\author[add2]{Guoxiang Huang}
\author[add6]{Chunbo Zhao}
\author[add6]{Wei Li}
\author[add6]{Tengfei Wu}
\author[add1,add2]{Guangqiang He\corref{cor1}}

% 声明标记系统（新增共同一作说明）
\cortext[cor1]{Corresponding author. \\ Email: gqhe@sjtu.edu.cn}
\fntext[t]{These authors contributed equally to this work.}

%% Abstract
\begin{abstract}
Lithium niobate (LN) microring resonators, characterized by an exceptionally high second-order nonlinear coefficient and superior electro-optic tunability, serve as an outstanding platform for the precise control of integrated quantum frequency combs (QFCs). In this study, we introduce a bipartite entanglement criterion to investigate the pairwise entanglement characteristics of QFCs generated via the spontaneous parametric down-conversion (SPDC) process in lithium niobate microring resonators operating below threshold. Furthermore, we propose a universal framework for analyzing multimode squeezing in quadratic frequency combs, enabling the realization of ultrabroadband and high-degree multimode squeezing. We further reveal the underlying physical mechanism: modulation instability (MI), regulated by temporal walk-off control, not only enables the formation of frequency combs but also induces multimode squeezing in the corresponding resonant modes. This study uncovers the previously unexplored role of on-chip multimode squeezing in quadratic frequency combs while facilitating collective noise suppression across multiple modes, thus holding substantial potential for advancing quantum precision measurement and quantum information processing.
\end{abstract}

\begin{graphicalabstract}
\includegraphics{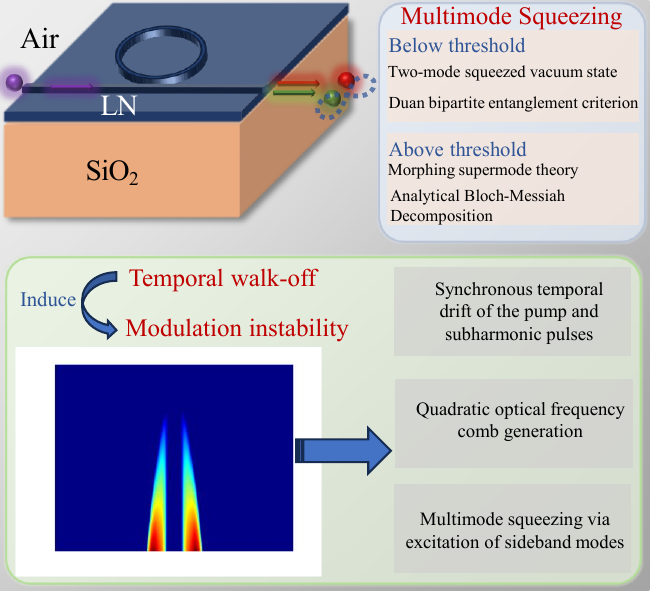}
\end{graphicalabstract}

\begin{highlights}
\item Lithium niobate microring quantum frequency combs achieve controllable entanglement 
\item Unified theory enables full-regime multimode squeezing optimization in OPO 
\item Modulation instability drives simultaneous comb-squeezing in quadratic systems 
\item Unveils quantum functionality for scalable information and precision metrology 
\end{highlights}
%% Keywords

\end{frontmatter}



\section{Introduction}
Squeezed optical states, as indispensable nonclassical resources in quantum information science, enable noise suppression beyond the standard quantum limit \cite{yap2020,suedbeck2020}, thereby facilitating advances in secure quantum communication and computation. The rapid progress in integrated photonics is catalyzing a shift toward chip-scale quantum optical systems \cite{yang2021}, where microring resonators have emerged as crucial platforms due to their strong optical confinement and enhanced nonlinear interactions \cite{ref1}.
Recent breakthroughs across diverse material platforms have significantly advanced microring-based quantum light sources. In silicon nitride (SiN) microrings, below-threshold spontaneous four-wave mixing has been employed to generate quadrature squeezed vacuum states and photon-number-difference squeezed states \cite{Wu2021IntegratedSM}, as well as to realize heralded single-photon sources \cite{PhysRevApplied.18.024059} and scalable multi-user quantum networks \cite{PhysRevA.109.040101}. Silicon carbide (SiC) microrings utilize soliton microcomb dynamics to achieve multimode entanglement \cite{guidry2022quantum}, while high-purity quantum sources have also been demonstrated in gallium nitride \cite{PhysRevLett.132.133603}, aluminum gallium arsenide \cite{PRXQuantum.2.010337}, and integrated nanophotonic platforms \cite{zhang2021}.
Complementing these advances, quantum frequency comb control techniques have evolved—from non-equilibrium driving of strongly coupled photonic dimer  systems \cite{tikan2021} to parity-time (PT) symmetric mode-selective pumping strategies \cite{PhysRevA.110.023714}—further underscoring the central role of quantum frequency combs in emerging applications such as quantum computing, communication, and precision metrology.

Lithium niobate (LN) offers distinct advantages over materials primarily governed by third-order ($\chi ^{(3)}$) nonlinearities, such as SiN, owing to its strong second-order ($\chi ^{(2)}$) nonlinear response \cite{shi2024efficient}. This intrinsic property enables direct exploitation of $\chi ^{(2)}$ processes, particularly SPDC, allowing for the efficient generation of entangled photon pairs with a significantly lower pump threshold than that required in SiN-based platforms \cite{PhysRevLett.124.163603,PhysRevLett.127.183601}.
In addition, LN facilitates quantum frequency conversion of single photons between the telecom band and the visible or mid-infrared range \cite{ma2019visible,doi:10.1126/science.aah5178}, enabling spectral interfacing across disparate quantum systems and meeting the stringent compatibility requirements of multimode quantum architectures. The heterogeneous integration of LN thin films with silicon-based photonic circuits \cite{churaev2023heterogeneously} further paves the way for multifunctional quantum photonic chips.
Moreover, lithium niobate exhibits a pronounced Pockels effect, enabling high-speed and low-loss electro-optic modulation, which plays a vital role in optical communication systems. Driven by these advantages, this work focuses on the theoretical investigation of multimode squeezing in quadratic frequency combs generated by LN microring resonators.

In this study, we develop an integrated LN microring resonator platform that enables the generation of QFCs through precise dispersion and coupling engineering \cite{li2024}. To elucidate the multimode squeezing mechanisms of quadratic frequency combs, we categorize our analysis into below-threshold and above-threshold regimes. Since below-threshold squeezing inherently occurs in a pairwise fashion between modes with quantum correlations in  quadratures, it gives rise to bipartite entanglement. We therefore employ a bipartite entanglement criterion to rigorously quantify the degree of entanglement between each mode pair. Our results show that under below-threshold conditions, increasing the pump power enables a controllable and continuous redshift of the squeezed frequency components. We further investigate the frequency-dependent squeezing characteristics and reveal the interaction between the detection frequency and the optimal readout phase \cite{xu2024frequency}.

Above the optical parametric oscillator (OPO) threshold, the coupled-mode equations for the signal-idler pair break down due to cascaded $\chi ^{(2)}$ nonlinear process. We therefore extend the model to a multimode formalism and derive quadratic coupled mean-field equations to describe the classical evolution of the optical frequency comb (OFC). Pronounced temporal walk-off in OPOs \cite{Hansson:18,Parra-Rivas:19} induces a novel MI band \cite{PhysRevA.93.043831}, wherein parametric amplification of the subharmonic field leads to enhanced pulse compression and synchronized temporal drift \cite{roy2022}.
As the classical mean-field framework fails to capture the quantum correlations underlying threshold behavior \cite{articleg}, we employ a supermode basis—quadrature-weighted combinations of frequency modes \cite{devalcarcel2006}—to characterize multimode squeezing and intermodal quantum correlations \cite{roslund2014}. In contrast to synchronously pumped OPOs (SPOPOs), where fluctuations are externally driven \cite{PhysRevResearch.5.023178}, quantum noise in our quadratic frequency comb arises intrinsically from $\chi ^{(2)}$ nonlinearity, linking the below-threshold quantum state to the mean-field solution.

During the detuning scan that culminates in soliton crystal formation, we observe pronounced quadrature squeezing, with ultrabroadband multimode correlations emerging between the pump and subharmonic modes. This phenomenon originates from the intricate interplay between optical nonlinearity and dispersion, wherein minor perturbations are exponentially amplified, yielding substantial MI gain at selective frequencies. While MI underpins the formation of optical frequency combs, it concurrently drives extensive multimode squeezing across resonant modes. In contrast to the multipartite entanglement analysis in Ref. \cite{jia2025}, our work provides a unified characterization of quadrature squeezing and spectral correlations across 800×2 modes that potentially contribute to the underlying multimode dynamics. We emphasize that a complete description must account for both pump and subharmonic field modes, as neglecting inter-band coupling can obscure critical squeezing phenomena. The general framework we develop for analyzing multimode squeezing in quadratic frequency combs is broadly applicable and holds strong potential for emerging applications such as high-throughput optical interconnects \cite{marin-palomo2017} and photonic machine learning accelerators \cite{xu2021}.

\section{\label{sec2}Microring Resonator Simulation Model}
Lithium niobate, renowned for its superior $\chi ^{(2)}$ nonlinearity, enables a high-frequency pump photon ($\Omega_p$) to spontaneously split into two lower-frequency photons—a process known as SPDC. These resultant photons, designated as the signal ($\Omega_s$) and idler ($\Omega_i$) photons, form an entangled pair. Notably, SPDC is exclusively initiated by  vacuum fluctuations and inherently adheres to both energy conservation and phase matching conditions.

In this study, we introduce an advanced on-chip microcavity architecture grounded in OPO theory, which ingeniously incorporates an enhanced coupling module to facilitate highly efficient nonlinear interactions. As depicted in Fig.~1(a), the micro-ring resonator comprises a ring waveguide coupled to a bus waveguide, with a z-cut LN ridge waveguide serving as the nonlinear medium. The waveguide is engineered with a silicon dioxide (SiO$_2$) lower cladding to suppress substrate leakage and enhance vertical optical confinement, while an air upper cladding is employed to further tailor the mode field distribution. The optical field circulates along the azimuthal direction of the micro-ring resonator, and the effective cross-sectional area ($A_{\mathrm{eff}}$) of the ring waveguide \cite{dudley2006supercontinuum}, quantified by Equation~(3), characterizes the extent of field localization within the resonator; a reduced effective mode area signifies superior confinement of the optical field.

\begin{equation}\label{Aeff} 
A_{\mathrm{eff}} = \frac{\left(\iint_{-\infty}^{+\infty} |F(y, z)|^{2} \, dy \, dz \right)^{2}}{\iint_{-\infty}^{+\infty} |F(y, z)|^{4} \, dy \, dz}. 
\end{equation} 
Here, $F(y,z)$ delineates the modal distribution within the LN and SiO$_2$ material system. In our theoretical framework, this distribution is presumed invariant along the longitudinal propagation direction within the resonator, remaining independent of the evolution of the optical field along the cavity. This assumption is instrumental in facilitating the application of the separation of variables technique for solving the coupled mode equations.

\begin{figure}[htb]
	\includegraphics[width=1\linewidth]{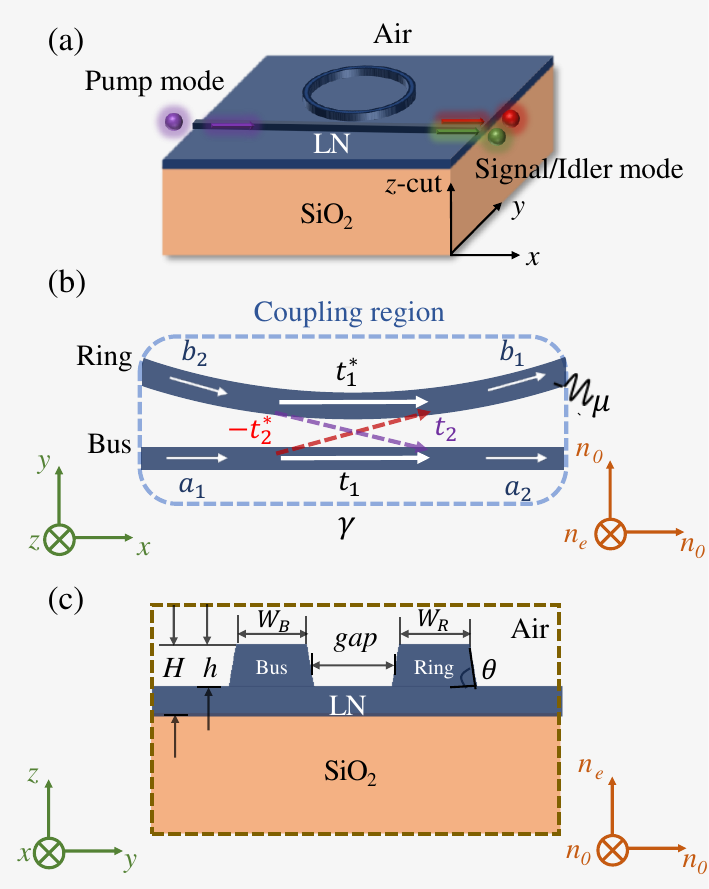}% Here is how to import EPS art
	\caption{\label{f1} (a)A three-dimensional rendering of the LN resonator is provided, with \( R \) denoting the mean radius of the ring waveguide.
(b) Coupling region between Ring and Bus with coordinate system and refractive indices indicated.
(c) Mode profile and refractive index distribution of a LN microring resonator.}
\end{figure}

The coupling region and the corresponding input–output schematic are illustrated in Fig.~1(b). The geometry of this region directly determines the coupling rate between the ring resonator and the bus waveguide, enabling precise control and forming a solid foundation for rigorous analysis of the resonator’s input–output characteristics. Figure~1(c) shows the cross-sectional profile of the ridge waveguide, with all relevant structural parameters clearly annotated. In this work, we focus on the electromagnetic properties of the fundamental transverse electric (TE) mode. The resonance profile of the cavity follows a Lorentzian lineshape. Due to the frequency dependence of the refractive index \( n(\omega) \), resonance peaks exhibit spectral shifts as a function of optical frequency. Accurate control over key cavity parameters—including detuning, dispersion, coupling strength, and loss rate—within the designed platform requires a comprehensive and physically consistent theoretical framework.

\subsection{Detuning and Dispersion }

This section delves into the fundamental mechanisms of detuning and dispersion. We introduce the relative mode number $l$ ($l \in \mathbb{Z}$) to index the modes adjacent to the half-harmonic mode $\omega_0 $ (corresponding to $l = 0$). By performing a Taylor expansion of the resonance mode frequency around $\omega_0$, we obtain:

\begin{equation}\label{Taylor}
\omega_l = \omega_0 + D_1 l + \frac{D_2}{2} l^2 + \cdots = \omega_0 + \sum_{k = 1}^{\infty} D_{k} \frac{l^k}{k!},
\end{equation}

where $D_1 = 2\pi \nu_f$, with $\nu_f$ representing the free spectral range (FSR). The parameter $D_2$ governs the group velocity dispersion ~\cite{shi2023}, where $D_2 > 0$ corresponds to anomalous dispersion, while $D_2 < 0$ characterizes normal dispersion. Higher-order dispersion terms ($D_k$ for $k \geq 3$) are neglected in this model. The integrated dispersion is defined as:

\begin{equation}
D_{\text{int}} = \omega_l - \omega_0 - D_1 l,
\end{equation}

which can be accurately approximated by a quadratic polynomial in the vicinity of $\omega_0$~\cite{godey2014}.

The cold-cavity pump detuning is given by:

\begin{equation}
\Delta_p = \omega_p - \Omega_p.
\end{equation}

Meanwhile, we set the detuning at the subharmonic mode to $\Delta=\omega_0-\Omega_0$. Here, $\Omega_l$ denote the comb frequency corresponding to the mode number $l$. Owing to $\Omega_p=2\Omega_0$, we obtain:
\begin{equation}
\Delta = (2\omega_0 - \omega_p+\Delta_p)/2.
\end{equation}

\subsection{\label{GAL}Coupling, Loss and Gap }

We analyze the fundamental relationship between coupling rate (\(\gamma\)), intrinsic loss rate (\(\mu\)), and resonator quality factor (\(Q\)). Following the virtual channel framework in Refs.~\cite{vernon2015,vernon2015b}, intrinsic losses are modeled as a phantom channel with beam-splitter-like behavior, under the constraint \(\gamma \ll \nu_f\), \(\mu \ll \nu_f\), where \(\nu_f\) denotes the free spectral range.

As illustrated in Fig.~1(b), the coupling and loss rates are given by:
\begin{equation}
\gamma = |t_2|^2 \nu_f = (1 - |t_1|^2)\nu_f, \quad
\mu = \alpha_c L \nu_f,
\end{equation}
where \(|t_1|^2\) and \(|t_2|^2\) are the energy reflection and transmission coefficients, respectively, and \(L = 2\pi R\) is the resonator perimeter with \(R = (D_{\text{in}} + D_{\text{out}})/4\). Here, $D_{\text{in}}$ and $ D_{\text{out}}$refer to the microring's inner and outer diameters, respectively. The absorption coefficient \(\alpha_c\) (in m\(^{-1}\)) is approximated as
\begin{equation}
\alpha_c \approx \frac{Q_0 R D_1}{\omega_0},
\end{equation}
with \(Q_0\) the intrinsic quality factor and \(\omega_0\) the resonance frequency.

To evaluate the effect of external coupling, we fix \(b_2 = 0\) and vary the ring–bus gap. The ratio \(r = \gamma/\mu\) defines the coupling regime: under-coupled (\(r < 1\)), critically coupled (\(r = 1\)), and over-coupled (\(r > 1\)). We focus on the over-coupled regime, which enables enhanced power extraction despite reduced intracavity field strength.

In LN resonators, the loaded quality factor governs field buildup and energy confinement:
\begin{equation}
Q = \omega_0 \tau_p = \frac{\omega_0}{\Gamma},
\end{equation}
where \(\tau_p\) is the photon lifetime and \(\Gamma\) the total linewidth. The intrinsic and coupling rates are
\begin{equation}
\mu = \frac{\omega_0}{Q_0}, \quad \gamma = \Gamma - \mu = \frac{\omega_0}{Q_{\mathrm{ex}}},
\end{equation}
with \(Q_{\mathrm{ex}}\) the external quality factor~\cite{javid2023}. The total \(Q\) satisfies
\begin{equation}
\frac{1}{Q} = \frac{1}{Q_0} + \frac{1}{Q_{\mathrm{ex}}},
\end{equation}
linking the coupling gap to \(Q_{\mathrm{ex}}\) and thereby enabling tunable external coupling.

\begin{figure}[htb]
	\includegraphics[width=1\linewidth]{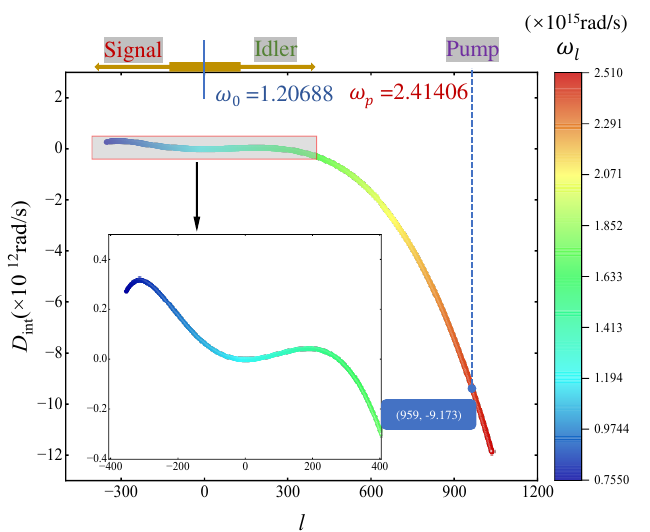}% Here is how to import EPS art
	\caption{\label{f3} Curve of Dispersion \( D_{\text{int}} \) vs. Mode Number \( l \). The diagram includes the regions of the signal and idle modes as well as the location of the pump mode.}
\end{figure}

\subsection{Actual Simulation}

Our objective is to engineer a dispersion-flattened waveguide by optimizing the structural parameters of the LN-integrated microcavity, such as its geometry and dimensions, to achieve phase matching in the SPDC process. This condition is fulfilled when the following equation holds:

\begin{equation}\label{pm}
	\Delta k = \frac{\omega_{p}n(\omega_{p})}{c}-\frac{\omega_{s}n(\omega_{s})}{c}-\frac{\omega_{i}n(\omega_{i})}{c}=0.
\end{equation}

Here, $\Delta k$ quantifies the phase mismatch. For our designed micro-ring resonator, at 1560 nm the first-order dispersion is given by
\(
k'_1 = \left.\frac{dk}{d\omega}\right|_{\omega_0} = 7879.9ps/m ,
\)
and the group velocity dispersion is
\(
k''_1 = \left.\frac{d^2k}{d\omega^2}\right|_{\omega_0} = -0.0219ps^2/m.
\)
Therefore, the FSR we have set can be expressed as: $\nu_f=1/(k'_1L)=201.976$GHz.
At 780 nm, the corresponding values are
\(
k'_2 = \left.\frac{dk}{d\omega}\right|_{\omega_p} = 8085.6ps/m,
\)
and
\(
k''_2 = \left.\frac{d^2k}{d\omega^2}\right|_{\omega_p} = 0.3624ps^2/m.
\)
We employ COMSOL Multiphysics to simulate the complete waveguide geometry and its integrated dispersion profile. A fifth-order polynomial fit is then applied to closely match the integrated dispersion \(D_{\text{int}}\) as depicted in Figure 2. Additionally, Lumerical FDTD simulations are conducted to elucidate the relationship between the gap and the coupling coefficient. 
Assuming an intrinsic quality factor of 
$Q_0 = 3.7 \times 10^{6},$
the intrinsic loss rate is computed as 
\(
\mu = 3.27 \times 10^{8}\ \text{rad/s}.
\)
With a coupling ratio 
\(
r = \frac{\gamma}{\mu} = 1.222,
\)
indicative of an over-coupled configuration, the external quality factor is determined to be 
\(
Q_{ex} = Q_0 / r = 3.03 \times 10^{6}.
\)
Based on the FDTD simulation outcomes, the gap is calculated to be 490 nm.

Figure 1(c) illustrates the structure of an LN microring resonator, featuring a radius of \( R = 100\,\mu\text{m} \), with both the waveguide and bus widths set at \( W_R = W_B = 2000\,\text{nm} \). The outer height is \( H = 600\,\text{nm} \), while the inner height is \( h = 410\,\text{nm} \), and the design incorporates an angle of \( \theta = 75^\circ \). Simulation results for this configuration indicate a pump mode frequency (\( \omega_p \)) of 2.41406×$10^{15}$\,rad/s, a fundamental mode (mode 0) frequency (\( \omega_0 \)) of 1.20688×$10^{15}$\,rad/s, and an effective area (\( A_{\text{eff}} \)) of \( 0.997\,\mu\text{m}^2 \).

\section{\label{sec5}Multimode Squeezing}
In this section, we present a comprehensive analysis of the multimode squeezing properties of quadratic frequency combs based on LN resonators. Below the OPO threshold, two-mode vacuum squeezing arises in correlated pairs, allowing the use of the Duan bipartite entanglement criterion to quantitatively evaluate the entanglement and squeezing of multiple two-mode squeezed states. Above the threshold, the evolution of the quadratic frequency comb is described by the quadratic-coupled mean-field equation. We further apply supermode theory in conjunction with the Analytical Bloch-Messiah Decomposition (ABMD) to characterize its multimode squeezing structure. Moreover, we identify that the MI gain driven by group-velocity mismatch plays a critical role in mediating the generation of multimode squeezing.
\subsection{Bipartite Entanglement Criterion}
The dynamics of the bipartite entangled system are detailed in Appendix A. To rigorously quantify the entanglement between the signal and idler modes, we employ the bipartite entanglement criterion from Ref.~\cite{PhysRevLett.84.2722} to evaluate the entanglement measure \( C_s \). The position (\(\hat{x}_j\)) and momentum (\(\hat{y}_j\)) quadrature operators for each mode  are elegantly formulated in terms of the annihilation and creation operators \(\hat{a}_j\) and \(\hat{a}_j^{\dagger}\):

\begin{equation}\label{bian}
\hat{x}_j = \frac{ 1}{\sqrt{2}}(\hat{a}_j^\dagger e^{\mathrm{i} \theta_{j}}+\hat{a}_j e^{-\mathrm{i} \theta_{j}}), \quad \hat{y}_j = \frac{ 1}{\sqrt{2}}(\mathrm{i}\hat{a}_j^\dagger e^{\mathrm{i} \theta_{j}}-\mathrm{i}\hat{a}_j e^{-\mathrm{i} \theta_{j}}).
\end{equation}

By adjusting the detection angles of the signal and idler beams (\(\theta_s, \theta_i\)), we derive:

\begin{equation}
	\left(\delta \hat{x}_{s}, \delta \hat{x}_{i}, \delta \hat{y}_{s}, \delta \hat{y}_{i}\right)^{\mathrm{T}}=P\left(\delta \hat{a}_{s}, \delta \hat{a}_{s}^{\dagger}, \delta \hat{a}_{i}, \delta \hat{a}_{i}^{\dagger}\right)^{\mathrm{T}}.
\end{equation}

Introduce the sum and subtraction basis:

\begin{equation}
	\hat{x}_{\pm}=\frac{1}{\sqrt{2}}(\hat{x}_{s} \pm \hat{x}_{i}), 
	\quad\hat{y}_{\pm}=\frac{1}{\sqrt{2}}(\hat{y}_{s} \pm \hat{y}_{i}),
\end{equation}
and the fluctuation vector is formulated as:
\begin{equation}
	\delta \hat{X}_{\pm}=\left(\delta \hat{y}_{+}, \delta \hat{x}_{+}, \delta \hat{y}_{-}, \delta \hat{x}_{-}\right)^{\mathrm{T}}
 =G\left(\delta \hat{x}_{s}, \delta \hat{x}_{i}, \delta \hat{y}_{s}, \delta \hat{y}_{i}\right)^{\mathrm{T}}.
\end{equation}

The spectral noise density matrix $S_{\hat{X}_{\pm}}(\omega)$ is determined as follows:
\begin{equation}
\begin{aligned}
		S_{\hat{X}_{\pm}}(\omega)&=\left\langle\delta \hat{X}_{\pm}(\omega) \delta \hat{X}_{\pm}^{\mathrm{T}}(-\omega)\right\rangle\\
		&=\frac{T_r\cdot S_{a}(\omega) \cdot T_r^{\mathrm{T}}+(T_r\cdot S_{a}(\omega) \cdot T_r^{\mathrm{T}})^{\mathrm{T}}}{2},
\end{aligned}
\end{equation}
where $T_r=G\cdot P=\frac{1}{2}
\begin{pmatrix}
- \mathrm{i} e^{-\mathrm{i} \theta_{s}} & \mathrm{i} e^{\mathrm{i} \theta_{s}} & -\mathrm{i} e^{-\mathrm{i} \theta_{i}} & \mathrm{i} e^{\mathrm{i} \theta_{i}} \\
e^{-\mathrm{i} \theta_{s}} & e^{\mathrm{i} \theta_{s}} & e^{-\mathrm{i} \theta_{i}} & e^{\mathrm{i} \theta_{i}} \\
- \mathrm{i} e^{-\mathrm{i} \theta_{s}} & \mathrm{i} e^{\mathrm{i} \theta_{s}} & \mathrm{i} e^{-\mathrm{i} \theta_{i}} & -\mathrm{i} e^{\mathrm{i} \theta_{i}} \\
e^{-\mathrm{i} \theta_{s}} & e^{\mathrm{i} \theta_{s}} & -e^{-\mathrm{i} \theta_{i}} & -e^{\mathrm{i} \theta_{i}}
\end{pmatrix}.$

The Duan criterion is expressed as follows \cite{Gonzalez-Arciniegas:17}:
\begin{equation}\label{criterion}
	C_s =((\Delta \hat{y}_{+})^{2}+(\Delta \hat{x}_{-})^{2}-|O|\geq 0,
\end{equation}
where $(\Delta \hat{x}_{-})^{2}=S_{\hat{X}_{\pm}}(\omega)(4,4)$,  $(\Delta \hat{y}_{+})^{2}=S_{\hat{X}_{\pm}}(\omega)(1,1)$, and $O=\cos(\theta_{s}-\theta_{i})$. If the Duan criterion is violated, i.e.,  
\(
C_s < 0,
\)
the bipartite modes are confirmed to be entangled. A lower value of \( C_s \) indicates a stronger degree of quantum entanglement.

\begin{figure}[htb]
	\includegraphics[width=1\linewidth]{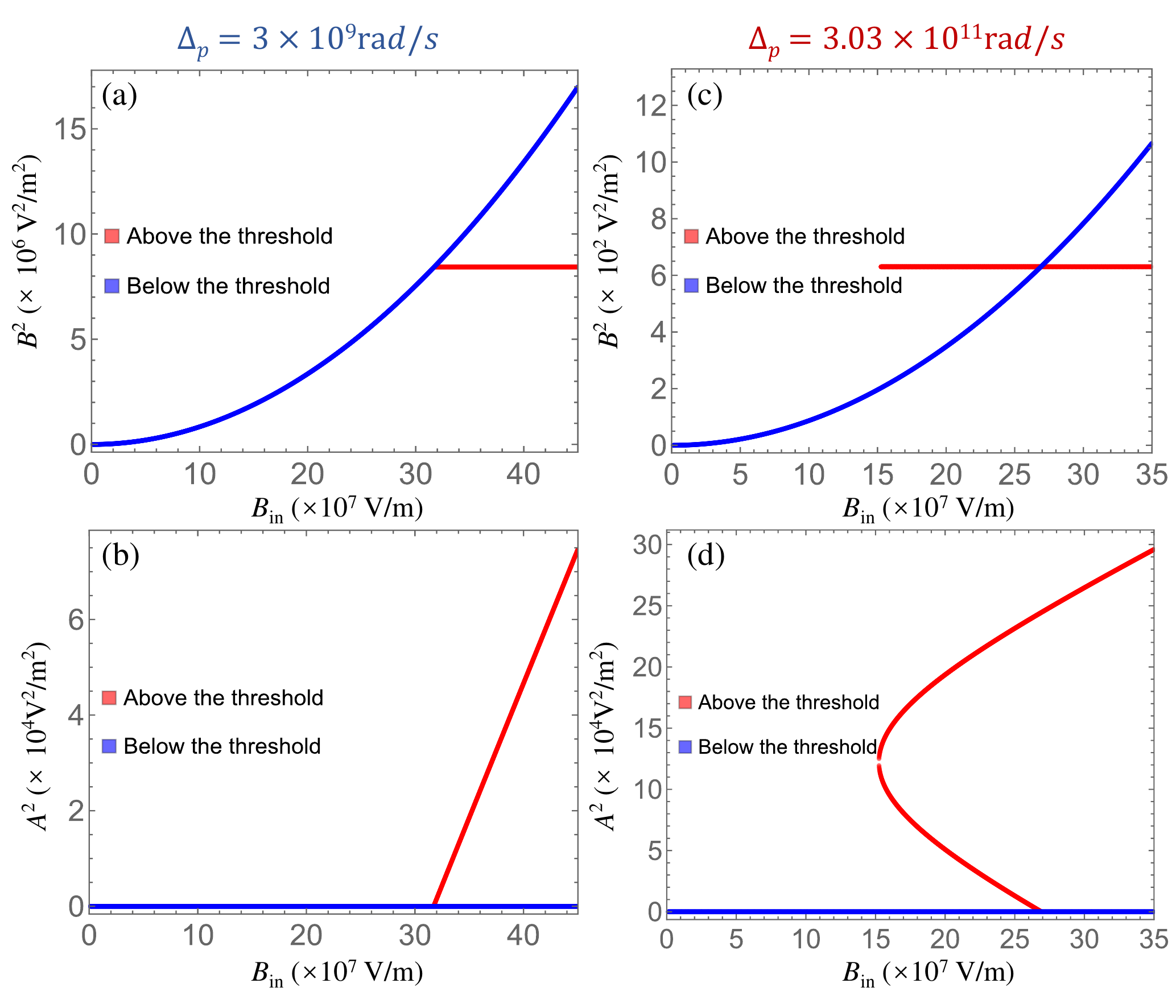}% Here is how to import EPS art
	\caption{\label{f5} Steady-state solutions for different pump detunings. \quad(a) Intracavity pump mode amplitude vs. injected pump mode amplitude (\(\Delta_ p=3 \times 10^9 \, \text{rad/s}\)). (b) Intracavity signal mode amplitude vs. injected pump mode amplitude (\(\Delta_ p = 3 \times 10^9 \, \text{rad/s}\)). (c) Intracavity pump mode amplitude vs. injected pump mode amplitude (\(\Delta_ p = 3.03 \times 10^{11} \, \text{rad/s}\)). (d) Intracavity signal mode amplitude vs. injected pump mode amplitude (\(\Delta_ p = 3.03 \times 10^{11} \, \text{rad/s}\)). }
\end{figure}
\subsection{Frequency-Dependent Squeezing}
Our platform inherently accommodates the entanglement analysis of multiple bipartite mode pairs. However, given the similarity of results across different modes, we focus our analysis on the first mode (\(l = 1\)) for conceptual clarity. Simulations are conducted at 20 °C, with thermal fluctuations disregarded. We compute the steady-state intracavity amplitudes of the pump and signal modes as functions of the injected pump amplitude using \textit{Mathematica} [Figs.~3(a), 3(b)]. The pump detuning is fixed at \(\Delta_p = 3 \times 10^9~\mathrm{rad/s}\). Below a critical threshold, the signal mode remains unexcited; above it, a nonzero amplitude emerges, defining the threshold power \(P_{\mathrm{th}}\).

Above threshold, the intracavity pump power saturates, while the signal power increases quasi-linearly. A markedly distinct behavior arises at large pump detuning (\(\Delta_p = 3.03 \times 10^{11}~\mathrm{rad/s}\)), where the subharmonic field acquires positive detuning (\(\Delta = 4.78 \times 10^8~\mathrm{rad/s}\)). Under these conditions, the signal power curve exhibits pronounced bending above threshold [Fig. 3(d)], while the intracavity pump power falls below the signal power—opposite to the behavior observed at low pump detuning.

Figure~4(a) plots \(P_{\mathrm{th}}\) as a function of mode number \(l\) under high detuning. At lower \(l\), the threshold increases with \(l\), indicating that modes further detuned from \(\omega_0\) are more readily excited—an inversion of the trend seen under weak detuning. This stems from modified phase-matching conditions, each dictating a distinct excitation threshold. Benefiting from lithium niobate’s high nonlinearity and favorable dispersion profile, the threshold power remains in the milliwatt regime—markedly lower than in conventional platforms.

We further implement a pump-tunable entanglement scheme. As shown in Fig.~4(b), increasing the pump amplitude induces a continuous redshift in the frequency maximizing entanglement. At threshold, this frequency approaches zero, accompanied by a substantial bandwidth broadening. The optimal frequency lies in the few-GHz domain, closely mirroring the multimode squeezing scenario discussed below. This spectral tunability highlights the potential for real-time optimization via pump power modulation.

Simulations with fixed \(\Delta_p = 3 \times 10^9~\mathrm{rad/s}\) and \(B_{\mathrm{in}} = 1.5 \times 10^8~\mathrm{V/m}\), and varying coupling strength \(r\), reveal analogous behavior [Fig.~4(c)]: progressive redshift of the optimal entanglement frequency, bandwidth broadening, and enhanced squeezing. This arises from overcoupling, which augments the SPDC gain and broadens the phase-matching bandwidth.

To capture spectral dependence, we define the readout angle \(\phi = \theta_s - \theta_i\). As shown in Fig.~4(d), the optimal readout angle varies monotonically with frequency, enabling its continuous adaptation. This frequency-dependent squeezing is highly pertinent for precision metrology, such as gravitational wave detection, offering a broadband, low-loss alternative to narrowband filtering cavities, and thus facilitating more efficient quantum-enhanced measurements.

\subsection{Quadratic Coupled Mean-Field Equation}
To investigate the frequency comb dynamics dominated by $\chi ^{(2)}$ nonlinearity, we now derive the quadratic coupled mean-field equations. Starting from the coupled-mode equations for the signal and idler modes, we extend them to a multimode framework while neglecting quantum fluctuations. We first derive the total Hamiltonian of the multimode system as follows:
\begin{figure}[htbp]
	\includegraphics[width=1\linewidth]{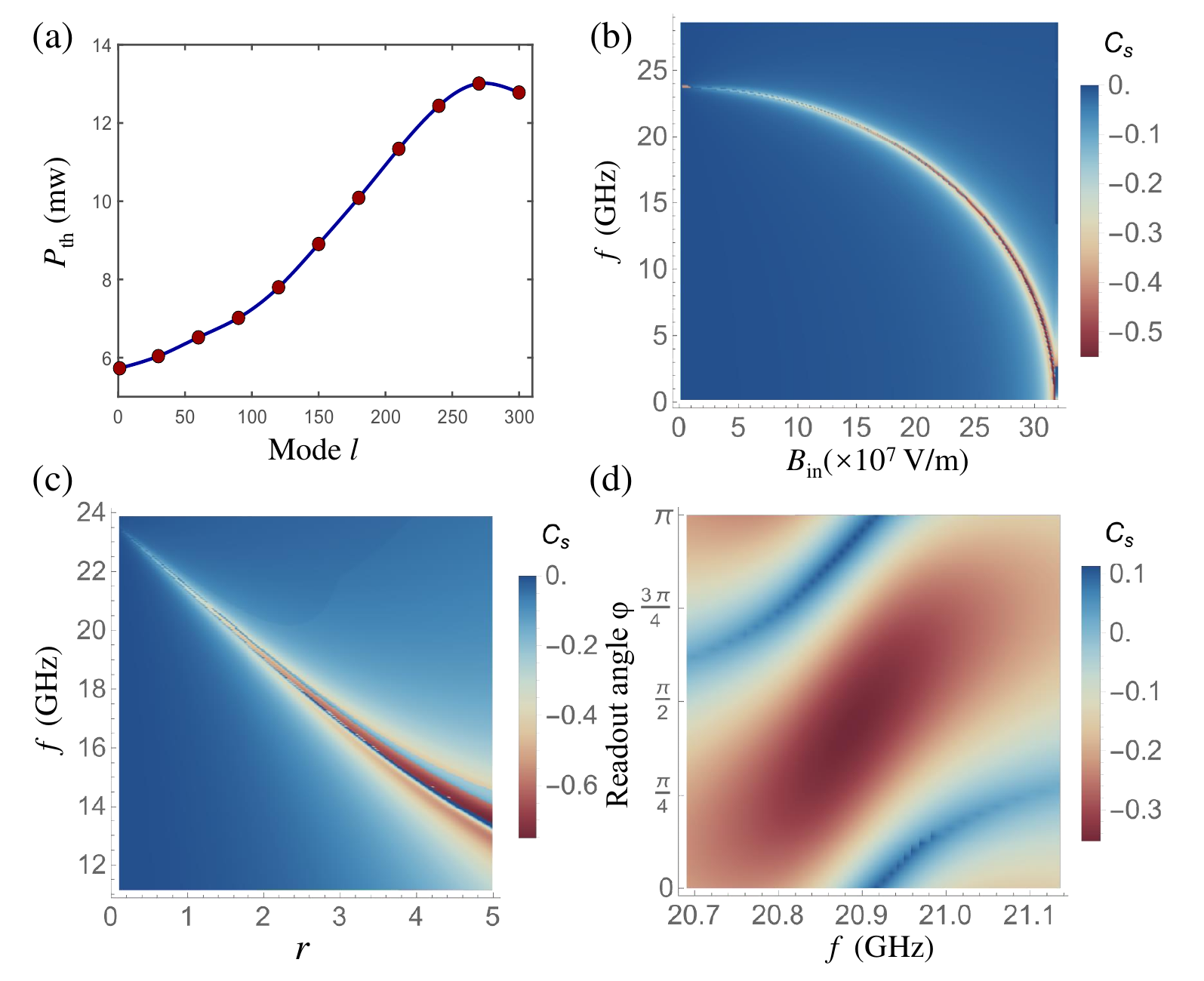}% Here is how to import EPS art
	\caption{\label{below}   (a) Pump threshold power for the corresponding mode number. (b) Relationship between observation frequency (Fourier frequency), injected pump mode amplitude, and entanglement degree. (c) Optimization of entanglement through tuning cavity coupling rate. (d) Relationship between readout angle, observation frequency, and entanglement degree which reflects frequency dependent squeezing.}
\end{figure}

\begin{figure*}[htbp]
	\includegraphics[width=1\linewidth]{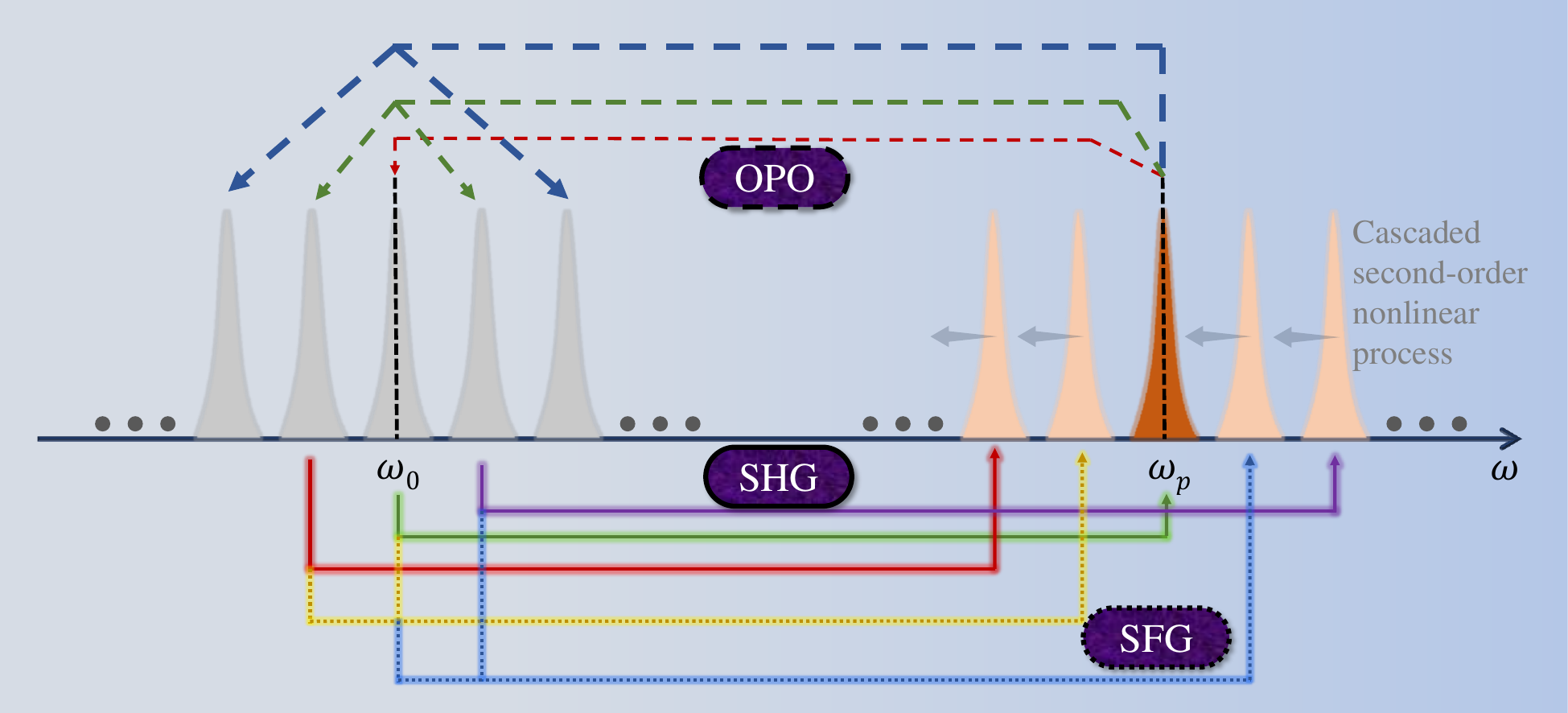}% Here is how to import EPS art
	\caption{\label{below}  Schematic of the initial steps in forming a  double OFC via second-order nonlinear processes. The top panel illustrates the down-conversion OPO process, while the bottom panel depicts the upconversion processes, including second harmonic generation (SHG) and sum-frequency generation (SFG). These processes are represented by dashed lines, solid lines, and dotted lines, respectively. }
\end{figure*}
\begin{equation}
\begin{aligned}
    \hat{H} &=\sum_{u}\hbar\omega_u \hat{a}_u^{\dagger}\hat{a}_u+\sum_{v}\hbar\omega_v \hat{b}_v^{\dagger}\hat{b}_v \\
    &-\frac{1}{2}\sum_{k,n,j}\hbar g_0 \left(\hat{a}_k\hat{a}_n\hat{b}_j^{\dagger}+\hat{a}_k ^{\dagger}\hat{a}_n^{\dagger}\hat{b}_j\right)\delta[\Omega_j-\Omega_k-\Omega_n],\\
    \end{aligned}
\end{equation}

where \( g_0 \) is the $\chi ^{(2)}$ nonlinear coefficient, assumed uniform for all modes.
This leads to the following coupled-mode equation for an arbitrary mode \( u \) of the subharmonic field and an arbitrary mode \( v \) of the pump field:
\begin{equation}\label{couple}
    \left\{
    \begin{aligned}
        \frac{\mathrm{d} \alpha_u} {\mathrm{d} t} &= \left[-\mathrm{i}(\omega_u-\Omega_u)-\Gamma \right]\alpha_u+\mathrm{i}g_{0}\sum_{k,j} \beta_j\alpha_k^{*}\delta[\Omega_j-\Omega_k-\Omega_u], \\
        \frac{\mathrm{d} \beta_v} {\mathrm{d} t} &= \left[-\mathrm{i}(\omega_v-\Omega_v)-\Gamma \right]\beta_v+\mathrm{i}g_{0}\sum_{k,n} \alpha_n\alpha_k\delta[\Omega_v-\Omega_k-\Omega_n]\\
        &+\sqrt{2\gamma}\beta^{in}\delta[\Omega_v-\Omega_p]. \\
    \end{aligned}
    \right.
\end{equation}
The range of \( u \) is defined as \( \left[ -\frac{N}{2} + 1, \frac{N}{2} \right] \), while the range of \( v \) is \( \left[ -\frac{N}{2} + 960, \frac{N}{2} + 959 \right] \). This choice ensures that the central mode number of the pump field is 959. Here, \( N \) represents the total number of modes considered for both the subharmonic and pump fields. The function \( \delta[\cdot] \) denotes the Dirac delta function, which takes a value of 1 when its argument is zero and 0 otherwise.

We define  
\(
\boldsymbol{\alpha} = \left\{ \alpha_{-N/2+1}, \dots, \alpha_{N/2} \right\}^{\text{T}}
\)  
and  
\(
\boldsymbol{\beta} = \left\{ \beta_{-N/2+960}, \dots, \beta_{N/2+959} \right\}^{\text{T}}
\),
allowing the system of \( 2N \) equations to be rewritten as two equations. Due to the computational complexity of the summation terms, we transform the frequency-domain convolution into a time-domain summation by applying the inverse discrete Fourier transform (IDFT) to the equations. We presume $a = \text{IDFT}[\boldsymbol{\alpha}], \quad b = \text{IDFT}[\boldsymbol{\beta}]$.
According to Ref. \cite{PhysRevA.93.043831}, the nonlinear effect compensates for the spectral phase induced by walk-off, allowing a LN resonator with significant walk-off to still be modeled using the mean-field approximation. Under this approximation, the evolution of the subharmonic and pump fields can be described by two coupled mean-field equations:
\begin{equation}\label{LLE}
    \left\{
    \begin{aligned}
        \frac{\mathrm{d} a} {\mathrm{d} t} &= \left(-\mathrm{i}\Delta-\Gamma \right)a-\mathrm{i} \frac{k''_1L \nu_f}{2} \frac{\partial^2 a}{\partial \tau^2}+\mathrm{i}g_{0}b\star a^{*}, \\
        \frac{\mathrm{d} b} {\mathrm{d} t} &= \left(-\mathrm{i}\Delta_{p}-\Gamma \right)b-\Delta k'L \nu_f \frac{\partial b}{\partial \tau}-\mathrm{i} \frac{k''_2L \nu_f}{2} \frac{\partial^2 b}{\partial \tau^2}\\
        &+\mathrm{i}g_{0}a\star a +\sqrt{2\gamma}B_{\text{in}}, \\
    \end{aligned}
    \right.
\end{equation}
where all parameters are consistent with those of the LN micro-ring resonator employed for QFC generation. Here, \(t\) denotes the cavity's evolution time, while the fast-time variable \(\tau\) spans the interval \(\left[-\frac{\tau_s}{2}, \frac{\tau_s}{2}\right]\), with \(\tau_s\) representing the round-trip time. The appearance of the second-order partial derivative with respect to \(\tau\) results from applying the IDFT to \(D_{\text{int}}\), while the first-order derivative term in \(\tau\) arises from the walk-off effect, which is induced by group velocity mismatch.
 We introduce the entrywise product operator \(\star\), which denotes element-wise multiplication between corresponding rows of column vectors \cite{guo2018}. Clearly, Equation (\ref{LLE}) admits a trivial all-zero solution for \( a \), whereas \( b \) does not exhibit such a trivial solution, corresponding to the sub-threshold region in Figure 3. Both trivial and nonzero solutions can exhibit MI gain, which amplifies random fluctuations, leading to the exponential growth of sidebands and ultimately forming a frequency comb \cite{PhysRevLett.121.093903}.
 By solving the mean-field equation using the split-step Fourier method, we can effectively simulate the dynamical evolution of the comb teeth.

 \begin{figure*}[htbp]
 \centering
	\includegraphics[width=0.65\linewidth]{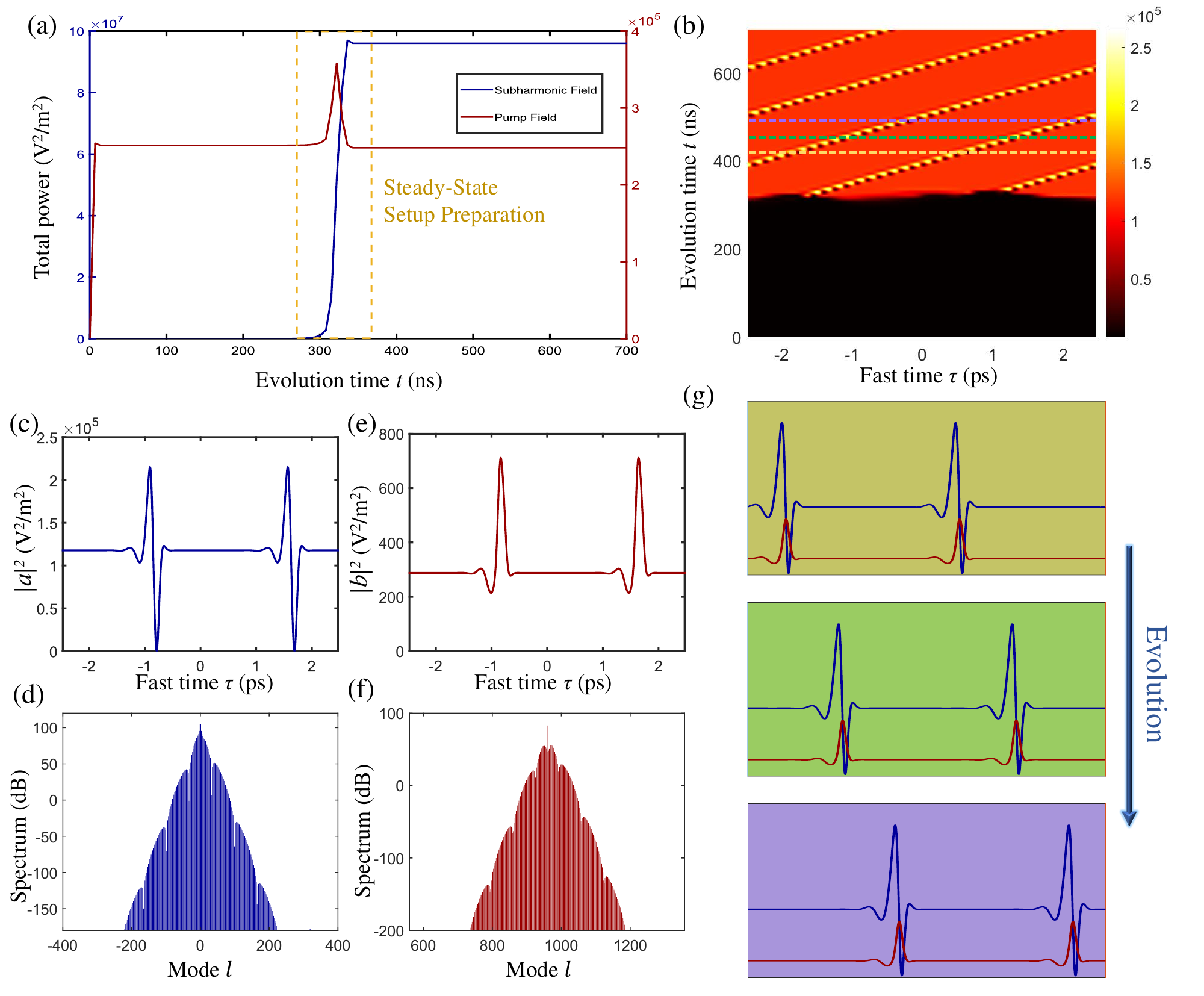}% Here is how to import EPS art
	\caption{\label{below1} (a) Evolution of the total power of the subharmonic and pump fields with respect to time \( t \). (b) Total power of the subharmonic field as a function of both evolution time and fast-time variable \( \tau \). (c) Time-domain pulse of the subharmonic field. (d) Corresponding subharmonic frequency spectrum, shown in terms of mode numbers. (e) Time-domain pulse of the pump field. (f) Corresponding pump frequency spectrum (with the center mode number at 959). (g) Cross-sections at different evolution times from panel (b), illustrating the pulse's collective drift behavior.
 }
\end{figure*}

\subsection{Modulation Instability-Induced  Frequency Comb}
In this subsection, we simulate the evolution of a quadratic OFC. To assist readers in comprehending the underlying mechanism, a schematic diagram illustrating the formation of the quadratic OFC is presented in Figure 5. The pump mode initiates the generation of multiple subharmonic modes via  OPO processes. These subharmonic modes subsequently regenerate pump components through second-harmonic generation (SHG) and sum-frequency generation (SFG). Facilitated by cavity-enhanced cascaded nonlinear interactions, this feedback loop drives the exponential proliferation of frequency components, ultimately resulting in the formation of a densely spaced, uniformly distributed quadratic OFC \cite{PhysRevA.91.063839}. We set $N = 800$, corresponding to the consideration of subharmonic field modes ranging from $-399$ to $400$ and pump field modes spanning from $560$ to $1359$, yielding a total of 1600 modes. Figure 6 presents the simulation outcomes at an evolution time of 700 ns, using a time step $dt = 0.1$ ps to ensure the robustness of the split-step Fourier method. To initiate the optical parametric oscillation process, random Gaussian white noise is seeded in each optical mode, with an injected pump amplitude of $B_{\text{in}} = 1.9 \times 10^8$ V/m.

Figure 6(a) delineates the temporal evolution of the aggregate power in both the subharmonic and pump fields. Initially, the subharmonic field exhibits negligible power due to the limited excitation of modes. As the evolution time reaches the marked interval in the figure, optical parametric amplification becomes prominent, leading to the excitation of a multitude of comb modes and a precipitous increase in total power. Simultaneously, the temporal gain-clipping mechanism is activated: due to group velocity mismatch, the signal pulse with the greatest temporal overlap with the pump pulse experiences the maximum gain, resulting in pulse compression. As the pump becomes depleted, gain saturation sets in, and the pulse centroid undergoes a shift driven by nonlinear acceleration \cite{roy2022}. Once spectral broadening and compression, as well as gain and loss, reach dynamic equilibrium—and temporal synchronization between the pump and subharmonic fields is established—the system ultimately settles into  steady state.

Figures 6(c) and 6(d) respectively exhibit the time-domain pulse and the OFC for the subharmonic field, while Figures 6(e) and 6(f) illustrate the corresponding time-domain pulse and OFC for the pump field. Under the stipulated simulation conditions, the time-domain profile reveals a bi-periodic pulse arrangement, corresponding to a spectral spacing of 2-FSR. It is noteworthy that, in contrast to soliton crystals, modulation instability combs are acutely sensitive to parameter variations, and the inter-pulse intervals are typically non-uniform. Extended simulation runs have corroborated the robust stability and coherence of the generated frequency comb.

\begin{figure*}[htbp]
\centering
	\includegraphics[width=0.7\linewidth]{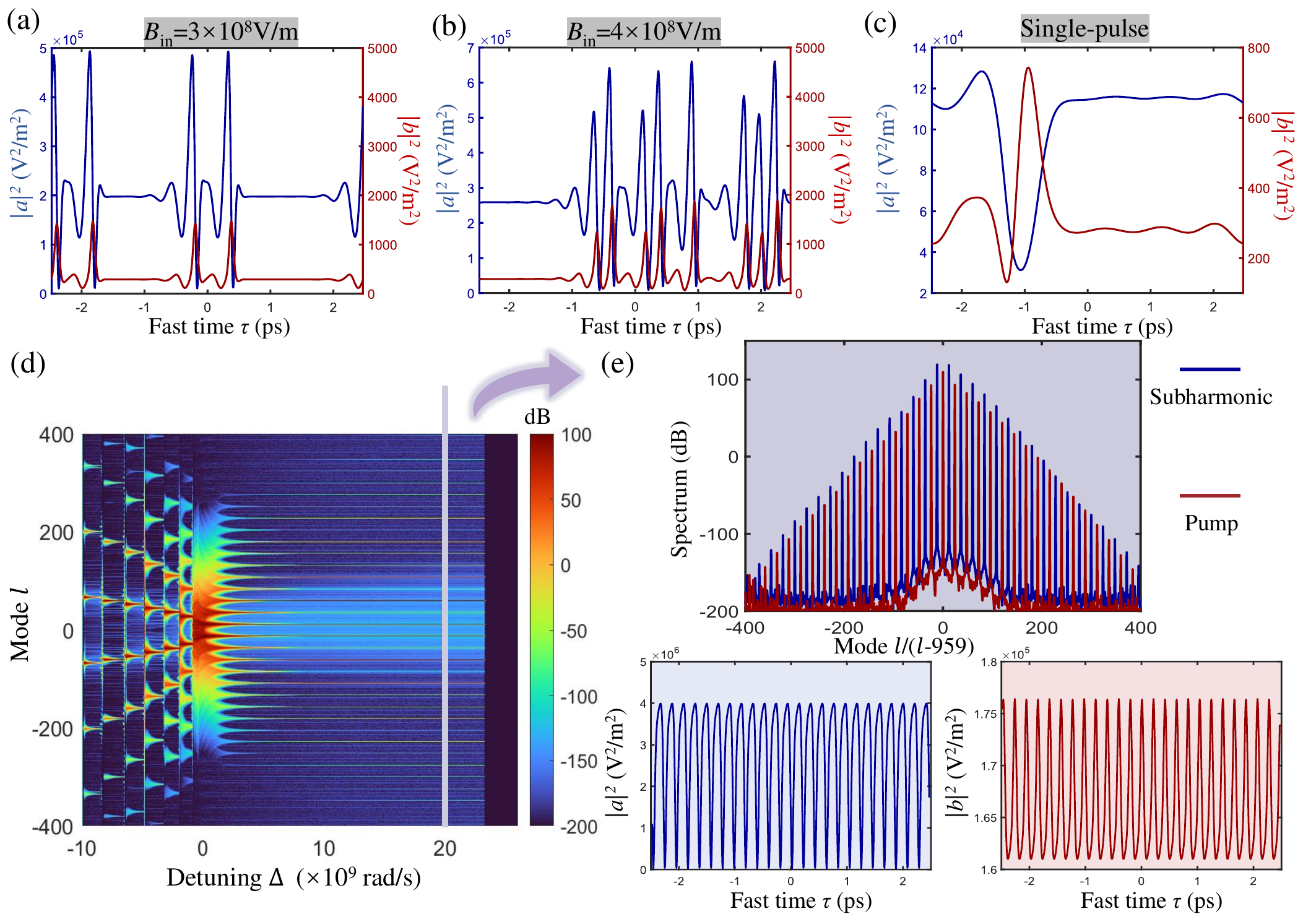}% Here is how to import EPS art
	\caption{\label{below2}  (a) Increase in the number of pulses with increasing injected pump amplitude. (b) Further increase in pulse count as the injected pump amplitude is increased. (c) Single-pulse scenario  under pump detuning sweep. (d) Evolution of the spectrum as a function of the detuning \( \Delta \).  
(e) Soliton crystal and the corresponding quadratic frequency comb at a detuning of \( \Delta = 2 \times 10^{10}~\mathrm{rad/s} \).
}
\end{figure*}
Figure 6(b) depicts the evolution of the subharmonic field intensity $|a(t,\tau)|^2$ as a function of the slow evolution time $t$ and the fast time variable $\tau$. The image reveals a series of evenly spaced, uniformly sloped “bright stripes”. To dissect this phenomenon more profoundly, we extracted three temporal cross-sections, as shown in Figure 6(g). Notably, as time progresses, the subharmonic and pump field pulses exhibit a synchronized drift at a constant velocity, a behavior consistent with the observations reported in Ref.~\cite{PhysRevA.93.043831}. The drift is significantly slower than the walk-off parameter \( \Delta k' \), indicating that the linear spectral phase induced by the walk-off effect is nonlinearly compensated. Fundamentally, this compensation enables group velocity locking between the subharmonic and pump waves \cite{PhysRevLett.116.033901,PhysRevLett.124.203902}.
This drift phenomenon manifests exclusively in the presence of walk-off and can be negated via a suitable transformation to a moving reference frame~\cite{jang2013}.

Furthermore, we observed that increasing the injected pump power leads to an augmentation in the number of pulses per period. For instance, when $B_{\text{in}} = 3 \times 10^8$ V/m, four pulses per period are observed; further increasing $B_{\text{in}}$ to $4 \times 10^8$ V/m results in the emergence of eight pulses, as illustrated in Figure 7. Our interpretation is twofold: first, a higher pump power accelerates gain saturation, thereby triggering competition among multiple pulses; second, intense nonlinear phase shifts accumulate within each pulse, inducing self-steepening in the time domain and ultimately causing the pulse to fragment into multiple sub-pulses~\cite{self}.
Moreover, although a single-pulse state can be accessed by tuning the detuning parameter, it merely corresponds to a saddle point—an unstable critical state—whereas multi-soliton states exhibit greater stability due to their more uniform energy distribution. By sweeping the detuning from \( -1 \times 10^{10}~\mathrm{rad/s} \) to \( 2.6 \times 10^{10}~\mathrm{rad/s} \), we obtain soliton crystal states, as illustrated in Figures~7(d) and 7(e). These states demonstrate excellent coherence and robustness against perturbations. The corresponding frequency combs are consistent with the nonlinear processes depicted in Figure~5, confirming the underlying formation mechanism.

\subsection{Supermode Theory}
To unveil the structural characteristics of the collective quantum fluctuations in the quadratic OFC, we extend the quantum fluctuation equations  to a multimode framework, yielding:
\begin{equation}\label{q}
    \left\{
    \begin{aligned}
        \frac{\mathrm{d} \delta \hat{a}_u} {\mathrm{d} t} &= \left[-\mathrm{i}(\omega_u-\Omega_u)-\Gamma \right]\delta\hat{a}_u+\sqrt{2\gamma}\delta \hat{a}_u^{\text{in}}+\sqrt{2\mu}\delta \hat{a}_u^{\text{loss}} \\
        &+\mathrm{i}g_{0}\sum_{k,j} (\alpha_k^{*}\delta \hat{b}_j+\beta_j \delta \hat{a}_k^{\dagger})\delta[\Omega_j-\Omega_k-\Omega_u],\\
        \frac{\mathrm{d} \delta \hat{b}_v} {\mathrm{d} t} &= \left[-\mathrm{i}(\omega_v-\Omega_v)-\Gamma \right]\delta\hat{b}_v+\mathrm{i}g_{0}\sum_{k,n} \alpha_k\delta \hat{a}_n\delta[\Omega_v-\Omega_k-\Omega_n]\\
        &+\sqrt{2\gamma}\delta \hat{b}_v^{\text{in}}+\sqrt{2\mu}\delta \hat{b}_v^{\text{loss}}. \\
    \end{aligned}
    \right.
\end{equation}
To transform the system of \( N \) coupled equations into a matrix form, we define the vectors:

\begin{equation}\label{d}
    \left\{
    \begin{aligned}
        \delta \hat{\mathbf{A}} &=
\begin{pmatrix}
\delta \hat{a}_{-N/2+1}, & \cdots & ,\delta \hat{a}_{N/2}
\end{pmatrix}, \\
        \delta \hat{\mathbf{B}} &=
\begin{pmatrix}
\delta \hat{b}_{-N/2+960}, & \cdots &, \delta \hat{b}_{N/2+959}
\end{pmatrix}, \\
\hat{\mathbf{\delta }}&=(\delta \hat{\mathbf{A}},\delta \hat{\mathbf{B}},\delta \hat{\mathbf{A}}^{\dagger},\delta \hat{\mathbf{B}}^{\dagger})^{\text{T}}.
    \end{aligned}
    \right.
\end{equation}
These vector representations enable us to express the quantum fluctuation dynamics in a compact matrix form as follows:
\begin{equation}\label{quantum}
	\frac{\mathrm{d}  \hat{\mathbf{\delta}}}{\mathrm{d} t}=M_{a} \cdot  \hat{\mathbf{\delta}}+U_{a}^{\mathrm{in}} \cdot \hat{\mathbf{\delta}}^{\mathrm{in}}+U_{a}^{\mathrm{loss}} \cdot  \hat{\mathbf{\delta}}^{\mathrm{loss}},
\end{equation}
\begin{figure*}[ht]  % [ht] 控制位置（here, top）
    \centering
\begin{equation}
M_a=\begin{pmatrix}
\chi_A & \mathrm{i}g_0NF_N^{-1}\text{diag}((F_N\alpha)^*)F_N & \mathrm{i}g_0NF_N^{-1}\text{diag}(F_N\beta)F_N^* & 0E_N \\
\mathrm{i}g_0NF_N^{-1}\text{diag}(F_N\alpha )F_N & \chi_B & 0E_N & 0E_N \\
-\mathrm{i}g_0NF_N^{-1}\text{diag}(F_N\beta ^*)F_N^* & 0E_N & \chi_A^* & -\mathrm{i}g_0NF_N^{-1}\text{diag}(F_N^*\alpha )F_N \\
0E_N& 0E_N & -\mathrm{i}g_0NF_N^{-1}\text{diag}(F_N\alpha^* )F_N &\chi_B^*
\end{pmatrix}.
\end{equation}
\end{figure*}
where $U_{a}^{\mathrm{in}}=\sqrt{2\gamma}E_{{4N}}$, $U_{a}^{\mathrm{loss}}=\sqrt{2\mu}E_{{4N}}$. To simplify the summation terms in the elements of the \( M_a \) matrix, we introduce the Fourier transform matrix \( F_N \), defined as
\(
F_N =
\frac{1}{N}
\begin{pmatrix}
W^0 & W^0  & \cdots & W^0 \\
W^0 & W^1  & \cdots & W^{N-1} \\
\vdots & \vdots  & \ddots & \vdots \\
W^0 & W^{N-1}  & \cdots & W^{(N-1)(N-1)}
\end{pmatrix}
\),
where \( W \) is given by
\(
W = e^{-\mathrm{i} \frac{2\pi}{N}}.
\)
By left-multiplying Equation (\ref{quantum}) with \( F_N \) and then subsequently left-multiplying by \( F_N^{-1} \), we can express \( M_a \) as Equation (24),
where $\chi_A=-\mathrm{i}\text{diag}(\Delta_{-N/2+1},\dots,\Delta_{N/2})-\Gamma E_N,\chi_B=-\mathrm{i}\text{diag}(\Delta_{-N/2+960},\dots,\Delta_{N/2+959})-\Gamma E_N$. Owing to $\hat{\delta}^{\text{out}}=-\hat{\delta}^{\text{in}}+U_a^{\text{in}}\hat{\delta}$, we can obtain $\hat{\delta}^{\text{out}}=M^{\text{in}}(\omega)\hat{\delta}^{\text{in}}+M^{\text{loss}}(\omega)\hat{\delta}^{\text{loss}}$, where $M^{\text{in}}(\omega)=U_a^{\text{in}}(\mathrm{i}\omega E_{4N}-M_a)^{-1}U_a^{\text{in}}- E_{4N}$.

To calculate the noise variance spectrum of any local oscillator, we define a supermode decay operator \cite{articleg}:
\begin{equation}
\hat{L}(t) = \sum_{l=-N/2+1}^{N/2} \eta_l \hat{a}_l(t)+\sum_{l=-N/2+960}^{N/2+959} \eta_l \hat{b}_l(t).
\end{equation}

Its Hermitian conjugate operator creates a superposition of photons between the longitudinal modes of the frequency comb, where
\(
\eta_l = |\eta_l| e^{\mathrm{i} \varphi_l}
\)
defines the amplitude and phase components of the supermode. After performing the transformation as in Equation (\ref{bian}), we can obtain the spectral noise density matrix:
\begin{equation}\label{sp}
S_x(\omega) = \frac{1}{\sqrt{2}}\left(\begin{matrix}
E_{2N} & E_{2N} \\
-\mathrm{i}E_{2N} & \mathrm{i}E_{2N}
\end{matrix} \right) \cdot M^\text{in}(\omega) \cdot \frac{1}{\sqrt{2}}\left( \begin{matrix}
E_{2N} & -\mathrm{i}E_{2N} \\
E_{2N} & \mathrm{i}E_{2N}
\end{matrix} \right).
\end{equation}
It is worth noting that, for computational convenience, the inherent cavity loss term has been excluded in Equation (\ref{sp}). This simplification is justified, as cavity losses only affect the extraction of the deformed supermodes, without impacting their composition. By applying ABMD, we can identify the maximum squeezing based on the given spectral noise density matrix. Combining the methods outlined in references \cite{PhysRevLett.125.103601} and \cite{HOUDE2024497}, we can implement the ABMD as follows: $S_x(\omega)=U(\omega)D(\omega)V^{\dagger}(\omega)$. Further details can be found in Appendix C.

In order to express the squeezing characteristics of cavity-deformed supermodes, we define the corresponding Hermitian quadrature operators as follows: 
\(
\hat{L}^{(+)}(t) = \frac{1}{\sqrt{2}} \left( \hat{L}^\dagger(t) + \hat{L}(t) \right), \hat{L}^{(-)}(t) = \frac{\mathrm{i}}{\sqrt{2}} \left( \hat{L}^\dagger(t) - \hat{L}(t) \right).
\)
Thus, the spectral noise of the quantum fluctuations can be expressed using the Wiener-Khinchin theorem \cite{PhysRevLett.125.103601}:

\begin{equation}
 \left\{
\begin{aligned}
V^+( \omega ) &= \int_{-\infty}^{\infty} \langle \hat{L}_{\text{out}}^{(+)}(t) \hat{L}_{\text{out}}^{(+)}(t+\tau) \rangle e^{-\mathrm{i}\omega \tau} d\tau,\\
V^-( \omega ) &= \int_{-\infty}^{\infty} \langle \hat{L}_{\text{out}}^{(-)}(t) \hat{L}_{\text{out}}^{(-)}(t+\tau) \rangle e^{-\mathrm{i}\omega \tau} d\tau .\\
\end{aligned}
\right.
\end{equation}

\begin{figure*}[htbp]
\centering
	\includegraphics[width=0.8\linewidth]{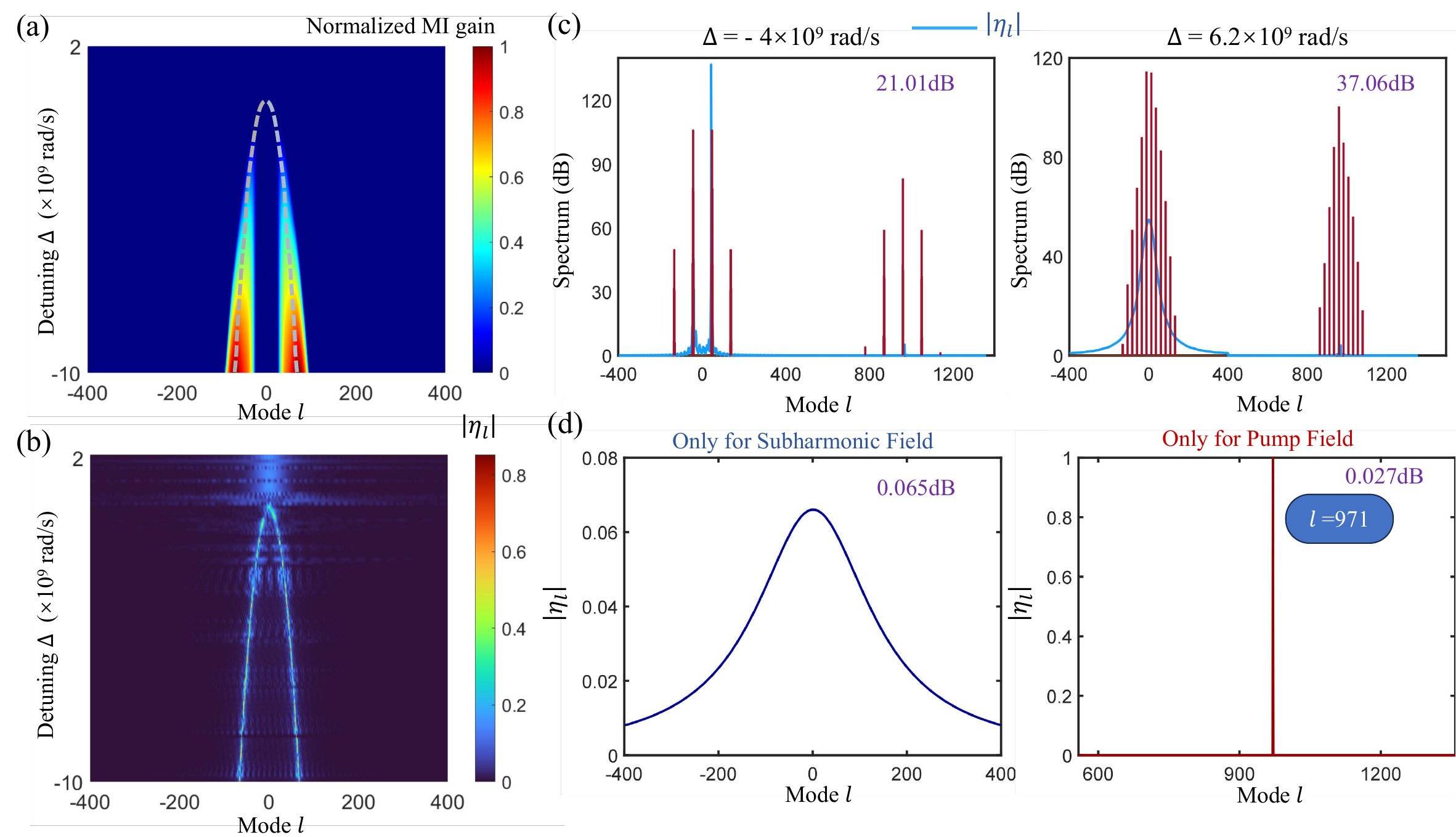}% Here is how to import EPS art
	\caption{\label{below}  (a) MI gain of the nontrivial solution induced by temporal walk-off. (b) Distribution of morphing supermode coefficients during the detuning sweep. (c) Spectra of quadratic frequency combs and the corresponding morphing supermode coefficient distributions at different detuning values. (d) Morphing supermode coefficients at \(\Delta = 6.2 \times 10^9~\mathrm{rad/s}\), considering only the subharmonic or pump field modes.
}
\end{figure*}
Here, \( V^+(\omega) \) and \( V^-(\omega) \) represent the squeezing or anti-squeezing of the cavity-deformed supermodes in the absence of intrinsic cavity loss.
The composition of the maximum squeezing cavity-deformed supermode corresponding to \( U(\omega) \) can be derived from the columns of \( U(\omega) \). When \( U(\omega) \) is a real matrix, the components of the \( k \)-th maximum squeezing cavity-deformed supermode satisfy:

\begin{equation}
 \left\{
\begin{aligned}
|\eta_{l,a}| &= \sqrt{|U(\omega)_{l,k}|^2 + |U(\omega)_{l+2N,k}|^2},\\
 \varphi_{l,a} &= \text{atan2}\left( U(\omega)_{l+2N,k}, U(\omega)_{l,k} \right),\\
|\eta_{l,b}| &= \sqrt{|U(\omega)_{l+N,k}|^2 + |U(\omega)_{l+3N,k}|^2},\\
\varphi_{l,b} &= \text{atan2}\left( U(\omega)_{l+3N,k}, U(\omega)_{l+N,k} \right).\\
\end{aligned}
\right.
\end{equation}
We can now elucidate the collective quantum fluctuation behavior in the context of cavity-deformed supermodes and their squeezing dynamics.

\subsection{Analysis of Multimode Squeezing}
By combining the quadratic optical frequency comb generated in Section~3.4 with the supermode formalism developed in Section~3.5, we employ the ABMD to  analyze the system’s multimode squeezing characteristics. To elucidate the underlying physical mechanism responsible for the emergence of multimode squeezing during the evolution of the quadratic frequency comb, we begin with the theoretical framework of MI gain. Specifically, MI yields both a trivial gain associated with the zero solution---corresponding to the unexcited parametric field---which is independent of group-velocity mismatch and determined solely by dispersion and pump strength, and a nontrivial MI gain driven by walk-off. The latter plays a critical role in the formation of the frequency comb. Details on MI theory and the gain of the trivial solution can be found in Appendix B.
Figure~8(a) shows the evolution of nontrivial MI gain as a function of cavity detuning, which directly corresponds to the distribution of the morphing supermode coefficients shown in Fig.~8(b), satisfying the normalization condition 
$\sum_l |\eta_l|^2 = 1.$
In this process, periodic modes far from the critical frequency become unstable, leading to their disappearance and the emergence of new modes closer to the MI gain maximum. As the peak MI gain frequency evolves with detuning~$\Delta$, previously stable sidebands eventually become sufficiently detuned from the critical frequency to become unstable, giving rise to new modes that align more closely with the evolving MI gain peak. This indicates that MI not only drives the formation of the quadratic frequency comb but also initiates the excitation of its corresponding squeezed modes.

Figure~8(c) displays the spectral profiles and supermode decomposition coefficients for both the Turing pattern and soliton crystal states. In both cases, strong multimode squeezing is observed, with squeezing levels reaching 21.01\,dB and 37.06\,dB, respectively. The supermode coefficients are predominantly concentrated around the subharmonic field modes, corresponding to their higher power occupancy. After the formation of the soliton crystal, the maximum of the supermode decomposition shifts to the central subharmonic mode, reflecting enhanced localization of squeezing in that mode.

To underscore the collective nature of multimode squeezing---where all modes within a spectral band must be incorporated to avoid the loss of quantum correlations due to inter-band coupling---we conduct a separate analysis of the subharmonic and pump fields. As illustrated in Fig.~8(d), the morphing supermode coefficients of the pump field are entirely concentrated in a single mode ($l = 971$), owing to minimal detuning and favorable phase-matching conditions.
Interestingly, the subharmonic field alone sustains significant squeezing, in some cases even exceeding the global squeezing level (see Fig.~C.1). In stark contrast, the pump field exhibits negligible squeezing when considered in isolation, rendering the multimode squeezing framework ineffective. This divergence stems from the fact that, even without subsequent SHG or SFG, the initial OPO process driven by the pump field can establish multimode squeezing in the subharmonic field. However, neglecting the modes involved in SHG and SFG disrupts the quantum nonlinear couplings essential for sustaining multimode entanglement. As a predominantly classical drive, the pump field itself lacks the capacity to generate quantum squeezing, and thus cannot form a coherent multimode squeezed system on its own.

\begin{figure}[htbp]
\centering
	\includegraphics[width=0.9\linewidth]{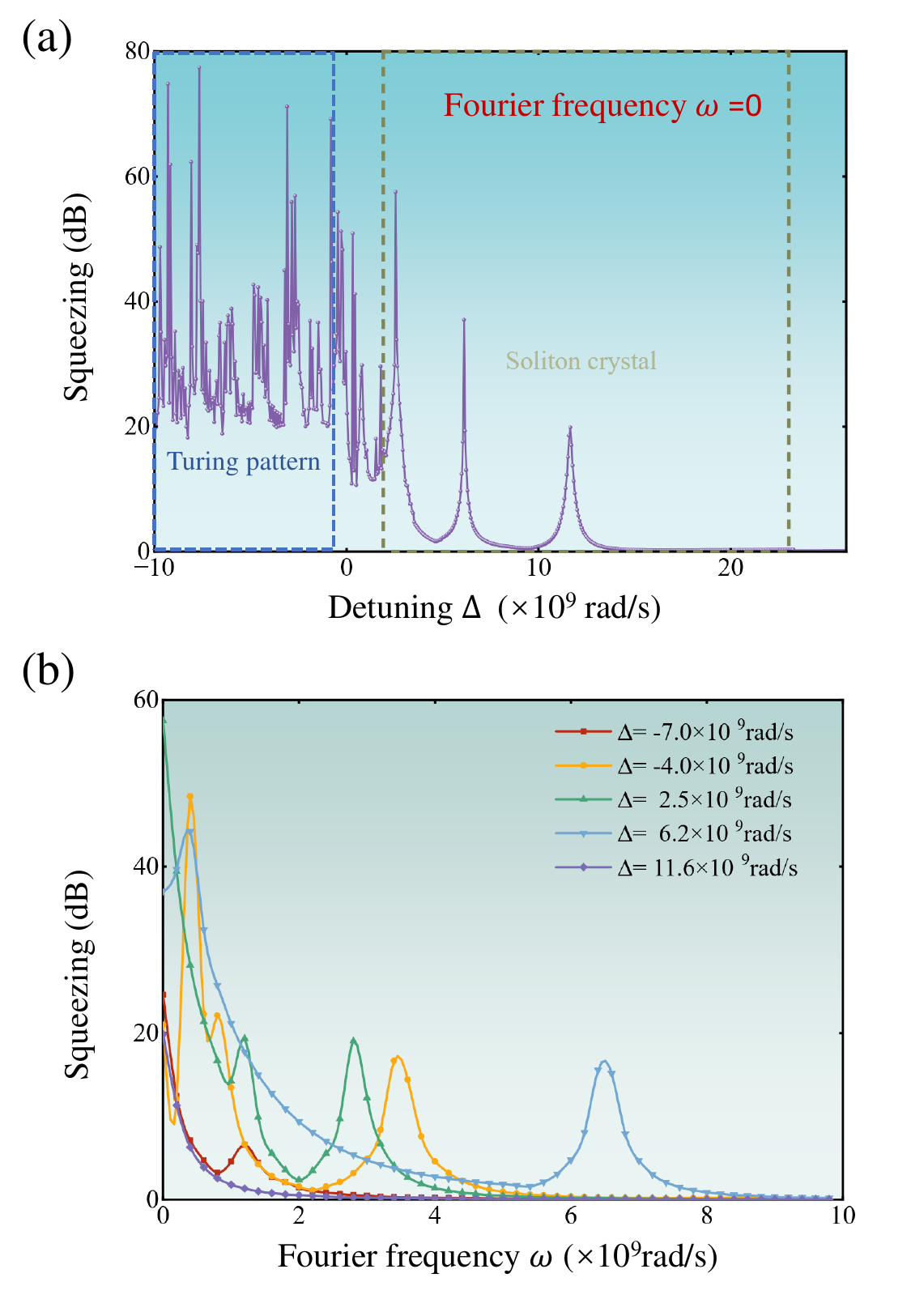}% Here is how to import EPS art
	\caption{\label{below}  (a) Evolution of multimode squeezing at zero Fourier frequency during the detuning sweep. (b) Spectral distribution of multimode squeezing at different detuning values.
}
\end{figure}
Figure 9(a) shows the evolution of multimode squeezing at zero Fourier frequency throughout the full detuning scan. Since the linearized fluctuation model is only valid on a stable mean-field background, our analysis focuses on the Turing pattern and soliton crystal regimes. In the Turing region, the degree of squeezing exhibits persistent oscillations, consistent with the periodic intracavity power variations during detuning. In contrast, in the soliton crystal regime, the squeezing level tracks the shift of the most strongly squeezed frequency with detuning. For large detuning, multimode squeezing vanishes entirely, indicating the breakdown of the soliton crystal structure.
Using the ABMD, we extract the spectral distribution of multimode squeezing, as shown in Fig. 9(b). The spectrum typically features two dominant peaks whose positions are highly sensitive to detuning. As detuning increases, only a single isolated squeezing peak remains near zero Fourier frequency. This behavior stands in stark contrast to pump-power tuning, where the squeezing peak positions remain nearly unchanged.

Multimode squeezed states, enabled by the ultrabroadband quantum noise suppression inherent to LN, significantly enhance the scalability and practicality of quantum technologies. These states allow for the generation of multimode squeezed light fields, enabling parallel access to quantum resources and supporting the advancement of distributed quantum computing. They also facilitate efficient implementation of multi-node quantum key distribution and quantum repeater protocols. In multiparameter quantum metrology, multimode squeezing offers substantial improvements beyond classical and standard quantum noise limits, with applications ranging from optical clock synchronization to biological imaging. Prospectively, as photonic chip integration continues to mature, this technology is poised to drive large-scale deployment in quantum networks and deep-space communication.

\section{\label{sec6}Conclusion}
We demonstrate a lithium niobate micro-ring platform that leverages $\chi ^{(2)}$ nonlinearity to generate entangled QFCs with milliwatt-level thresholds. Through dispersion and coupling engineering, we achieve tunable frequency dependent squeezing and quantify entanglement across modes. Leveraging the quadratic  coupled mean-field equation, we conduct an in-depth exploration of the evolutionary dynamics of multimode frequency combs. A supermode analysis reveals ultra-broadband quadrature squeezing, essential for scalable quantum networks and parallel quantum processing. We further propose and elucidate a physical mechanism wherein MI, beyond triggering the emergence of quadratic frequency combs, also drives multimode quadrature squeezing across the relevant modes.This work establishes LN as a superior material for integrated quantum photonics, bridging nonlinear dynamics and quantum optics for chip-scale applications.

\medskip
\textbf{Acknowledgements} \par %delete if not applicable))
This work was supported by the National Natural Science Foundation of China (Grant No. 62075129), Microwave Photonics Technology Key Laboratory of Sichuan Province (Grant No.2023-04) and the Science and Technology on Metrology and Calibration Laboratory (Grant No. JLKG2024001B002). 

\appendix
\section{Bipartite Entanglement Dynamics}
This section delves into the quantum dynamics of the resonator and presents a detailed analysis of two-mode quadrature squeezing.
\subsection{System Hamiltonian}
The system Hamiltonian rigorously characterizes the optical nonlinear dynamics in the quantum regime, with the resonant  modes described by the annihilation operators $\hat{a}_s$, $\hat{a}_i$, and $\hat{b}$. Here, $\hat{a}_s$ and $\hat{a}_i$ correspond to the signal and idler modes in the 1560 nm band, while $\hat{b}$ denotes the pump mode in the 780 nm band.

\setcounter{figure}{0}  % 重置图计数器为0

The free Hamiltonian $\hat{H}_0$ is given by:

\begin{equation}
\hat{H}_0 = \hbar \left(\omega_s \hat{a}_s^\dagger \hat{a}_s+\omega_i \hat{a}_i^\dagger \hat{a}_i+\omega_p \hat{b}^\dagger \hat{b}\right).
\end{equation}

For SPDC, the nonlinear Hamiltonian is formulated as:
\begin{equation}
\hat{H}_{\text{NL}} =- \hbar g_0\left(\hat{a}_s\hat{a}_i\hat{b}^\dagger+\hat{a}_s^\dagger\hat{a}_i^\dagger\hat{b}\right).
\end{equation}

Here, \( g_0 \) represents the second-order nonlinear coefficient of LN and can be expressed as \cite{PhysRevApplied.23.024030}:
\begin{equation}
g_0 = 2 \epsilon_0 \chi^{(2)} \sqrt{\frac{\hbar \omega_s \omega_i \omega_p}{16\pi \epsilon_0^3\epsilon_1^3  A_{\text{eff}} R},}
\end{equation}
where \(\chi^{(2)}\) denotes the second-order nonlinear susceptibility tensor of LN,  
\(\epsilon_0\) and \(\epsilon_1\) represent the vacuum permittivity and relative permittivity, respectively,  
\(A_{\text{eff}}\) is the effective mode area, and \(R\) denotes the radius of the microcavity.
 For simplicity, higher-order nonlinear effects beyond second-order nonlinearity are neglected.

The total Hamiltonian can be expressed as:
\begin{equation}
\hat{H} = \hat{H}_0 + \hat{H}_{NL}.
\end{equation}

\subsection{Heisenberg-Langevin Equations}
The Heisenberg-Langevin equation \cite{PhysRevA.93.063820} elegantly unifies the Heisenberg equation of motion with Langevin noise terms, providing a comprehensive description of the dynamical evolution of an open quantum system influenced by quantum fluctuations. This framework can be succinctly formulated as follows, with the signal mode serving as an illustrative example:
\begin{equation}
\left\{
\begin{aligned}
\frac{d \hat{a}_s}{dt}& = \frac{\mathrm{i}}{\hbar} [\hat{a}_s, \hat{H}] + V, \\
 V & = -\Gamma\hat{a}_s+\sqrt{2\gamma}\hat{a}_s^{\text{in}}+\sqrt{2\mu}\hat{a}_s^{\text{loss}}.
\end{aligned}
\right.
\end{equation}
\begin{figure*}[htb]
    \centering
    \includegraphics[width=0.7\textwidth]{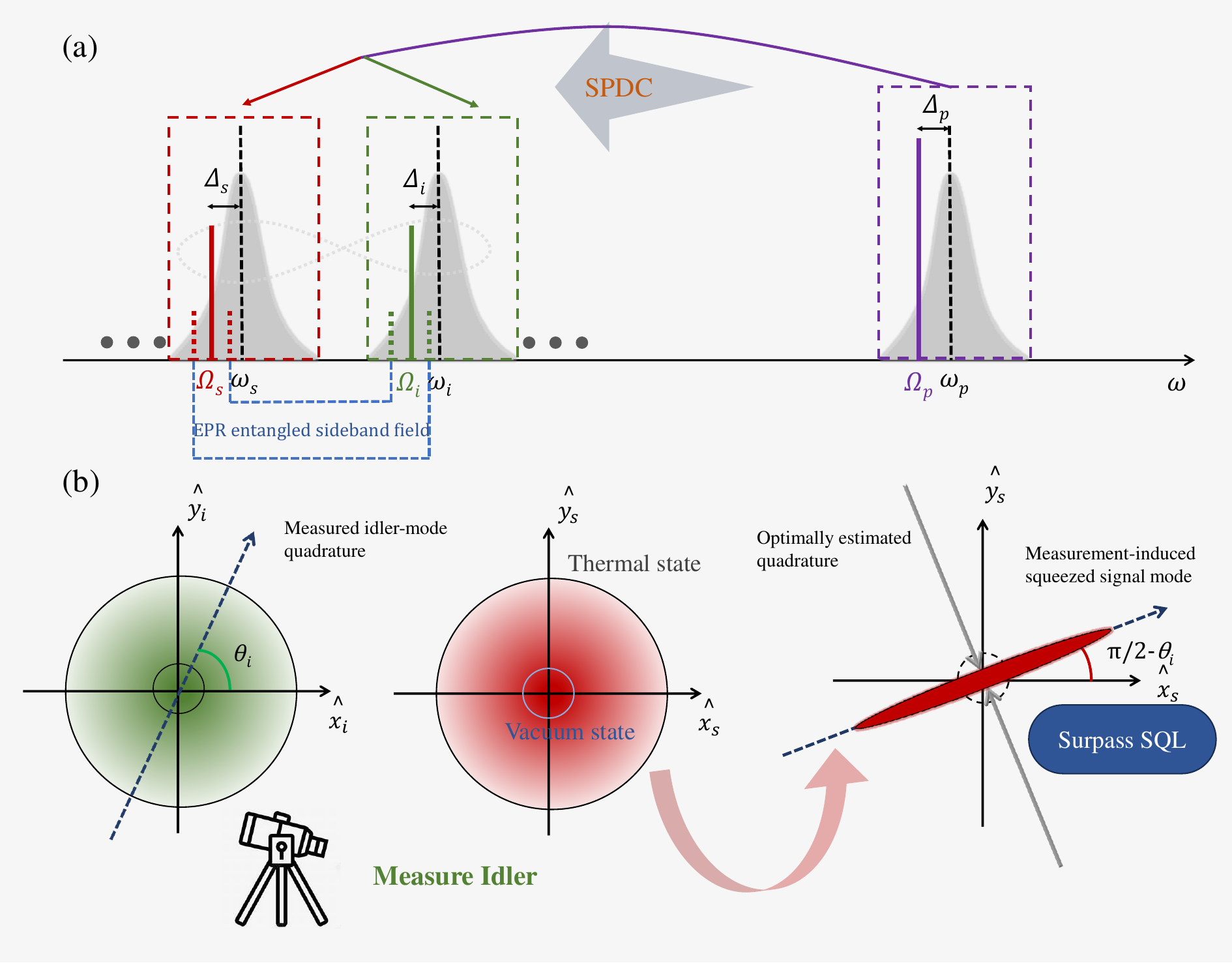}
    \caption{ (a) SPDC process in a resonant cavity and the formation of EPR entangled sidebands. (b) Quantum statistics of idle and signal modes and frequency-dependent squeezing.}
\end{figure*}

Here, the considered modes are presumed to exhibit analogous field distributions and are governed by a unified total loss rate $\Gamma$, which encapsulates the intrinsic dissipation rate $\mu$ and the coupling rate $\gamma$, satisfying the relation $\Gamma = \gamma + \mu$. The annihilation operators $\hat{a}^{\text{in}}$ and $\hat{a}^{\text{loss}}$ respectively represent the input and loss modes of the resonator. Within this theoretical construct, the loss mode is postulated to reside in the vacuum state, while the incident signal and idler modes are likewise assumed to be in vacuum states. The expectation value of the input pump mode is articulated as:

\begin{equation}
\left\langle\hat{b}^{\mathrm{in}}(t)\right\rangle = B_{\text{in}} = \sqrt{\frac{P_{\text{in}}}{\hbar \Omega_p}},
\end{equation}
where $P_{\mathrm{in}}$ (Watt) is the pump laser power.

By invoking the rotating wave approximation, wherein $\hat{a}_j e^{-i\Omega_j t}$ replaces $\hat{a}_j$ $(j=s,i)$ and $\hat{b} e^{-i\Omega_p t}$ replaces $\hat{b}$, the Heisenberg-Langevin equations governing the dynamical evolution of the pump, signal, and idler modes are elegantly expressed as:

\begin{equation}\label{Heisenberg-Langevin equation}
    \left\{
    \begin{aligned}
        \frac{\mathrm{d} \hat{b}} {\mathrm{d} t} &= \left(-\mathrm{i}\Delta_{p}-\Gamma \right)\hat{b}+\mathrm{i}g_{0}\hat{a}_{s}\hat{a}_{i}+\sqrt{2\gamma}\hat{b}^{\mathrm{in}}+\sqrt{2\mu}\hat{b}^{\mathrm{loss}}, \\
       \frac{\mathrm{d} \hat{a}_{s}} {\mathrm{d} t} &= \left(-\mathrm{i}\Delta_{s}-\Gamma \right)\hat{a}_{s}+\mathrm{i}g_{0}\hat{a}_{i}^{\dagger}\hat{b}+\sqrt{2\gamma}\hat{a}_{s}^{\mathrm{in}}+\sqrt{2\mu}\hat{a}_{s}^{\mathrm{loss}}, \\
       \frac{\mathrm{d} \hat{a}_{i}} {\mathrm{d} t} &= \left(-\mathrm{i}\Delta_{i}-\Gamma \right)\hat{a}_{i}+\mathrm{i}g_{0}\hat{a}_{s}^{\dagger}\hat{b}+\sqrt{2\gamma}\hat{a}_{i}^{\mathrm{in}}+\sqrt{2\mu}\hat{a}_{i}^{\mathrm{loss}}, \\
    \end{aligned}
    \right.
\end{equation}
where  \( \Delta_{s} \) and \( \Delta_{i} \) represent the normalized cold cavity detunings corresponding to the signal and idler modes, respectively. Since $\Omega_p = \Omega_s + \Omega_i$, the phase terms in the rotating wave approximation mutually cancel, simplifying the equations.

\subsection{Steady-State Equations}
We implement a linearization procedure by decomposing each field operator $\hat{a}_j$ into its steady-state mean value $\alpha_j$ and a fluctuation term $\delta \hat{a}_j$, expressed as
$
\hat{a}_j = \alpha_j + \delta \hat{a}_j
$
(The pump mode is set as $\hat{b} = \beta + \delta \hat{b}$).

Under steady-state conditions, $\alpha_j$ remains constant, allowing us to impose $\delta \hat{a}_j = 0$ and
$
\frac{d\alpha_j}{dt} = 0,
$
thereby deriving the steady-state Heisenberg-Langevin equations. Within this framework, the input fields of the signal and idler modes, along with the loss modes of the pump, signal, and idler, are all assumed to be in the vacuum state, satisfying:
\begin{equation}
\alpha_{s}^{\mathrm{in}}=\alpha_{i}^{\mathrm{in}}=\alpha_{s}^{\mathrm{loss}}=\alpha_{i}^{\mathrm{loss}}=\beta^{\mathrm{loss}}=0.
\end{equation}
Without loss of generality, we take the phase of the external pump as the reference, yielding the following definitions:
 $\alpha_{j}=A_{j} \mathrm{e}^{\mathrm{i}\theta_{j}}$, $\beta=B \mathrm{e}^{\mathrm{i}\theta_{\mathrm{p}}}$, $\beta^{\mathrm{in}}=B_{\mathrm{in}} \mathrm{e}^{\mathrm{i}\theta_{\mathrm{in}}}$. 
For  simplicity, we set $A_{s}=A_{i}=A$ and $\Delta_{s}=\Delta_{i}=\Delta$. 
Utilizing these parameters, we obtain the following expressions:

\begin{align}\label{steady-state solution}
    \left\{
    \begin{array}{ll}
         \left[\Delta^2+\Gamma^2-(g_0B)^2\right]A^2 = 0,  \\
         2\gamma B^2B_{\mathrm{in}}^2 = (\Delta_pB^2-\Delta A^2)^2+(\Gamma B^2+\Gamma A^2)^2.  \\
    \end{array}
    \right.
\end{align}
By solving Equation (\ref{steady-state solution}), we derive the steady-state relationship among the input pump \( B_{\text{in}} \), \( B \), and \( A \).

\begin{figure*}[htb]	\includegraphics[width=0.8\linewidth]{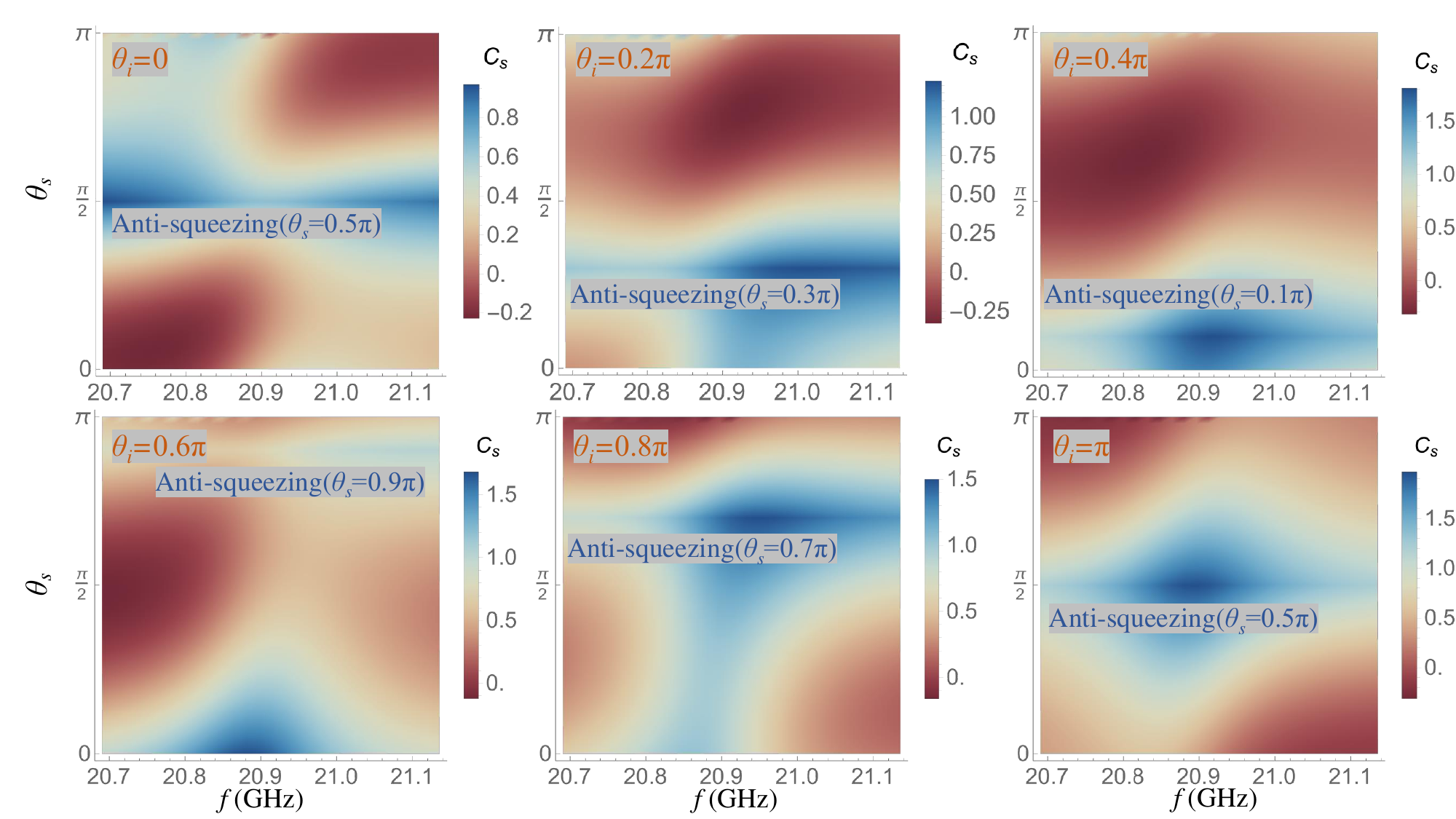}% Here is how to import EPS art
\centering
	\caption{\label{below3} The relationship between the entanglement degree, \( \theta_s \), and observation frequency is shown, with a fixed measurement angle \( \theta_i \) for the idle modes ($\Delta_p=3 \times10^9$rad/s,$B_{\text{in}}=1.5 \times10^8$V/m). The figure highlights the region of inverse squeezing corresponding to \( \theta_s \).
 }
\end{figure*}

\subsection{Quantum Fluctuations}
To rigorously examine the quantum properties of the signal and idler modes, we derive the quantum fluctuation equations from Eq.~(\ref{Heisenberg-Langevin equation}). Given that the steady-state solutions have been systematically determined, these fluctuation equations are obtained by subtracting the steady-state components from Eq.~(\ref{Heisenberg-Langevin equation}). In this framework, the pump field is treated as a classical parameter, thereby neglecting fluctuations in the pump mode (\(\delta \hat{b} = 0\)), while higher-order fluctuation terms are disregarded for analytical tractability.

We formulate the fluctuation vector for the signal and idler modes as follows:
\begin{equation}
	\delta \hat{\mathbf{A}}=\left(\delta \hat{a}_{s} \mathrm{e}^{-\mathrm{i} \theta_{s}}, \delta \hat{a}_{s}^{\dagger} \mathrm{e}^{\mathrm{i} \theta_{s}}, \delta \hat{a}_{i} \mathrm{e}^{-\mathrm{i} \theta_{i}}, \delta \hat{a}_{i}^{\dagger} \mathrm{e}^{\mathrm{i} \theta_{i}}\right)^{\mathrm{T}},
\end{equation}
where \(\theta_j\) represents the phase of the steady-state mean value, expressed as  
$
\alpha_j = A_j e^{i\theta_j}.
$
The time evolution of the fluctuations \(\delta \hat{a}_j\) is dictated by the following set of linearized equations:
\begin{equation}
	\frac{\mathrm{d} \delta \hat{\mathbf{A}}}{\mathrm{d} t}=M_{a} \cdot \delta \hat{\mathbf{A}}+U_{a}^{\mathrm{in}} \cdot \delta \hat{\mathbf{A}}^{\mathrm{in}}+U_{a}^{\mathrm{loss}} \cdot \delta \hat{\mathbf{A}}^{\mathrm{loss}},
\end{equation}
where $U_{a}^{\mathrm{in}}=\mathrm{diag}\left(\sqrt{2\gamma}, \sqrt{2\gamma}, \sqrt{2\gamma}, \sqrt{2\gamma}\right)$, $U_{a}^{\mathrm{loss}}=\mathrm{diag}\left(\sqrt{2\mu}, \sqrt{2\mu}, \sqrt{2\mu}, \sqrt{2\mu}\right)$. The matrix \( M_a \) arises from the linearization process, with its elements dictated by the mean field amplitudes and detuning parameters. Below the threshold (i.e., \( A = 0 \)), \( M_a \) can be expressed as:
\begin{equation}
M_a=\begin{pmatrix}
-\mathrm{i}\Delta_s-\Gamma & 0 & 0 & \mathrm{i}g_0\frac{\sqrt{2\gamma}B_{\mathrm{in}}}{\mathrm{i}\Delta_p+\Gamma} \\
0 & \mathrm{i}\Delta_s-\Gamma & \mathrm{i}g_0\frac{\sqrt{2\gamma}B_{\mathrm{in}}}{\mathrm{i}\Delta_p-\Gamma} & 0 \\
0 & \mathrm{i}g_0\frac{\sqrt{2\gamma}B_{\mathrm{in}}}{\mathrm{i}\Delta_p+\Gamma} & -\mathrm{i}\Delta_i-\Gamma & 0 \\
\mathrm{i}g_0\frac{\sqrt{2\gamma}B_{\mathrm{in}}}{\mathrm{i}\Delta_p-\Gamma}& 0 & 0 &\mathrm{i}\Delta_i-\Gamma
\end{pmatrix}.
\end{equation}

The frequency-domain dynamics of these fluctuations are obtained via a Fourier transform. Incorporating this transformation with the resonator’s input-output relations,  
$
\hat{a}^{\text{out}} = -\hat{a}^{\text{in}} + \sqrt{2\gamma} \hat{a},
$
provides a comprehensive characterization of the system’s response:

\begin{equation}
	\begin{aligned}
		\delta \hat{\mathbf{A}}^{\mathrm{out}}(\omega) &=-\delta \hat{\mathbf{A}}^{\mathrm{in}}+U_{a}^{\mathrm{in}} \delta \hat{\mathbf{A}}\\
		&=[U_{a}^{\mathrm{in}}\left(\mathrm{i} \omega E-M_{a}\right)^{-1} U_{a}^{\mathrm{in}}-E] \cdot \delta \hat{\mathbf{A}}^{\mathrm{in}}\\
  &+U_{a}^{\mathrm{in}}\left(\mathrm{i} \omega E-M_{a}\right)^{-1} U_{a}^{\mathrm{loss}}\cdot \delta \hat{\mathbf{A}}^{\mathrm{loss}},
	\end{aligned}
\end{equation}
where $E$ is the identity matrix.

Thus, the output spectral noise density matrix is formulated as:
\begin{equation}
	\begin{aligned}
		S_{a}(\omega)&=\left\langle\delta \hat{\mathbf{A}}^{\mathrm{out}}(\omega) \delta \hat{\mathbf{A}}^{\mathrm{out} ,\mathrm{T}}(-\omega)\right\rangle \\
		&=[U_{a}^{\mathrm{in}}\left(\mathrm{i} \omega E-M_{a}\right)^{-1} U_{a}^{\mathrm{in}}-E]\cdot M_c \\ 
  &\cdot [U_{a}^{\mathrm{in}}\left(\mathrm{-i} \omega E-M_{a}\right)^{-1} U_{a}^{\mathrm{in}}-E]^T
		+U_{a}^{\mathrm{in}}\left(\mathrm{i} \omega E-M_{a}\right)^{-1}\\
  &\cdot U_{a}^{\mathrm{loss}}\cdot M_c\cdot [U_{a}^{\mathrm{in}}\left(\mathrm{-i} \omega E-M_{a}\right)^{-1}U_{a}^{\mathrm{loss}}]^T,
	\end{aligned}
\end{equation}
where the matrix $M_c=\left(\begin{array}{cccc} 0 & 1 & 0 & 0 \\0 & 0 & 0 & 0 \\0 & 0 & 0 & 1 \\0 & 0 & 0 & 0\end{array}\right)$.

\subsection{Two-Mode Quadrature Squeezing }
The Duan  criterion \cite{PhysRevLett.84.2722,Gonzalez-Arciniegas:17} enables a precise quantitative characterization of pairwise entanglement and quadrature squeezing in the sub-threshold regime. Figure A.1(a) illustrates the process of forming multiple pairs of entangled states via spontaneous parametric down-conversion (SPDC) below the oscillation threshold in a LN microring resonator OPO. As shown in Figure A.1(b), when measuring the idle frequency mode at a specific angle $\theta_i$, the signal mode exhibits a squeezing effect in the $-\theta_i$ direction, with statistical properties superior to the vacuum state, thereby surpassing the standard quantum limit. In contrast, the signal mode demonstrates anti-squeezing along the orthogonal direction, $\pi/2 - \theta_i$. This phenomenon is further validated in Figure A.2, where we observe that for a given $\theta_i$, the maximum value of positive entanglement (corresponding to anti-squeezing) occurs at $\theta_s = \pi/2 - \theta_i$ or $\theta_s = 3\pi/2 - \theta_i$.

\section{Modulation Instability Gain}
\setcounter{figure}{0} 
The dynamics of the OPO can be described by an infinite-dimensional map of the field amplitudes, which, as shown in Ref.~\cite{PhysRevLett.121.093903,PhysRevLett.116.033901}, can be reduced to a single mean-field equation governing the subharmonic field \( a \):

\begin{equation}
\begin{aligned}
&\frac{\mathrm{d} a(t, \tau)}{\mathrm{d} t} =\,
 \left( -\Gamma - \mathrm{i}\Delta \right) a -\mathrm{i} \frac{k''_1L \nu_f}{2} \frac{\partial^2 a}{\partial \tau^2}\\
&+ \mathrm{i}\frac{\sqrt{2\gamma}g_0B_{\mathrm{in}}}{\nu_f}a^{*} - \frac{g_0^2}{\nu_f} a^{*} \star\left[ (a \star a) \otimes I(\tau) \right] .\\
\end{aligned}
\end{equation}
Here, we assume the phase-matching condition is satisfied and $\otimes$ denotes the convolution operation. The temporal response function is given by 
\( I(\tau) = \mathrm{IDFT}[ \hat{I}(\Omega)] \), where
\begin{equation}
\hat{I}(\Omega) = \frac{1 - \mathrm{i}x - e^{-\mathrm{i}x}}{x^2}, \quad 
x = -\Delta k' L \Omega - \frac{1}{2}k_2'' L \Omega^2.
\end{equation}
The frequency \( \Omega \) denotes the offset
angular frequencyrelative to \( \omega_0 \).

It is straightforward to verify that the mean-field equation admits a time-independent constant solution of the form 
\( a_0(t, \tau) = |a_0| e^{\mathrm{i}\phi} \). 
When \( |a_0| = 0 \), this corresponds to the trivial steady-state solution of the below-threshold OPO. 
Substituting this ansatz into the mean-field equation, we obtain the following condition:
\begin{equation}
- (\Gamma + \mathrm{i}\Delta) e^{2\mathrm{i}\phi} 
- \frac{g_0^2 \hat{I}(0)}{\nu_f} |a_0|^2 e^{2\mathrm{i}\phi} 
+ \mathrm{i}\frac{\sqrt{2\gamma }g_0}{\nu_f} B_{\mathrm{in}} = 0.
\end{equation}
Solving the above equation yields the steady-state intensity of the subharmonic field:
\begin{equation}\label{18}
|a_0|^2 = \frac{-\Gamma \nu_f \pm \sqrt{2\gamma g_0^2 B_{\mathrm{in}}^2 - \Delta^2\nu_f^2}}{g_0^2 \hat{I}(0)}.
\end{equation}

To analyze the stability of the constant solution against the growth of new frequency components, we perform a modulation instability (MI) analysis. Specifically, we consider the following ansatz for the subharmonic field:
\begin{equation}
a(t, \tau) = a_0 + A_1 e^{\mathrm{i}\Omega \tau} + A_2 e^{-\mathrm{i}\Omega \tau},
\end{equation}
where \( A_1 \) and \( A_2 \) represent small perturbations around the steady-state amplitude \( a_0 \). Substituting this ansatz into the mean-field equation and retaining only the first-order terms in \( A_1 \) and \( A_2 \), we project onto the corresponding frequency components. This yields the following set of linearized differential equations governing the evolution of the three field amplitudes:
\begin{equation}
\left\{
\begin{aligned}
\dot{a}_0 &= -(\Gamma + \mathrm{i}\Delta) a_0 - \frac{g_0^2}{\nu_f} |a_0|^2 a_0\, \hat{I}(0) + \mathrm{i}\frac{\sqrt{2\gamma}g_0}{\nu_f} B_{\mathrm{in}} a_0^*,  \\
\dot{A}_1 &= -\left[ \Gamma + \mathrm{i}(\Delta - \frac{k''_1L\nu_f}{2} \Omega^2) + \frac{2g_0^2}{\nu_f} |a_0|^2 \hat{I}(-\Omega) \right] A_1\\
&- \left[ \frac{g_0^2}{\nu_f}a_0^2 \hat{I}(0) - \mathrm{i}\frac{\sqrt{2\gamma}g_0}{\nu_f} B_{\mathrm{in}} \right] A_2^*,  \\
\dot{A}_2 &= -\left[ \Gamma + \mathrm{i}(\Delta - \frac{k''_1L\nu_f}{2} \Omega^2) + \frac{2g_0^2}{\nu_f} |a_0|^2 \hat{I}(\Omega) \right] A_2\\
&- \left[ \frac{g_0^2}{\nu_f}a_0^2 \hat{I}(0) - \mathrm{i}\frac{\sqrt{2\gamma}g_0}{\nu_f} B_{\mathrm{in}} \right] A_1^*. 
\end{aligned}
\right.
\end{equation}
The coupled evolution of the perturbations can be recast into a compact matrix form as:
\begin{equation}
\begin{pmatrix}
\dot{A}_1^* \\
\dot{A}_2
\end{pmatrix}
=
\mathbf{M}
\begin{pmatrix}
A_1^* \\
A_2
\end{pmatrix}.
\end{equation}

For the nontrivial steady-state solution, the eigenvalues of the matrix $\mathbf{M}$ are given by:
\begin{equation}
\begin{aligned}
\lambda_{\pm} &= -\left[\Gamma + \frac{g_0^2}{\nu_f} |a_0|^2 \iota_+(\Omega)\right] \\
&\pm \sqrt{ \left(\Gamma^2 + \Delta^2\right) - \left[\Delta - \frac{k''_1L\nu_f}{2} \Omega^2 - \mathrm{i}\frac{g_0^2}{\nu_f} |a_0|^2 \iota_-(\Omega) \right]^2 },
\end{aligned}
\end{equation}
where the field amplitude $|a_0|$ is related to the input field $B_{\mathrm{in}}$ via Eq.~(\ref{18}), and 
$\iota_{\pm}(\Omega) = \hat{I}(\Omega) \pm \hat{I}^{*}(-\Omega).$

The nontrivial constant solution loses stability  when the MI gain satisfies $\mathrm{Re}[\lambda_+] > 0$.
\begin{figure}[htb]
   \centering
    \includegraphics[width=0.5\textwidth]{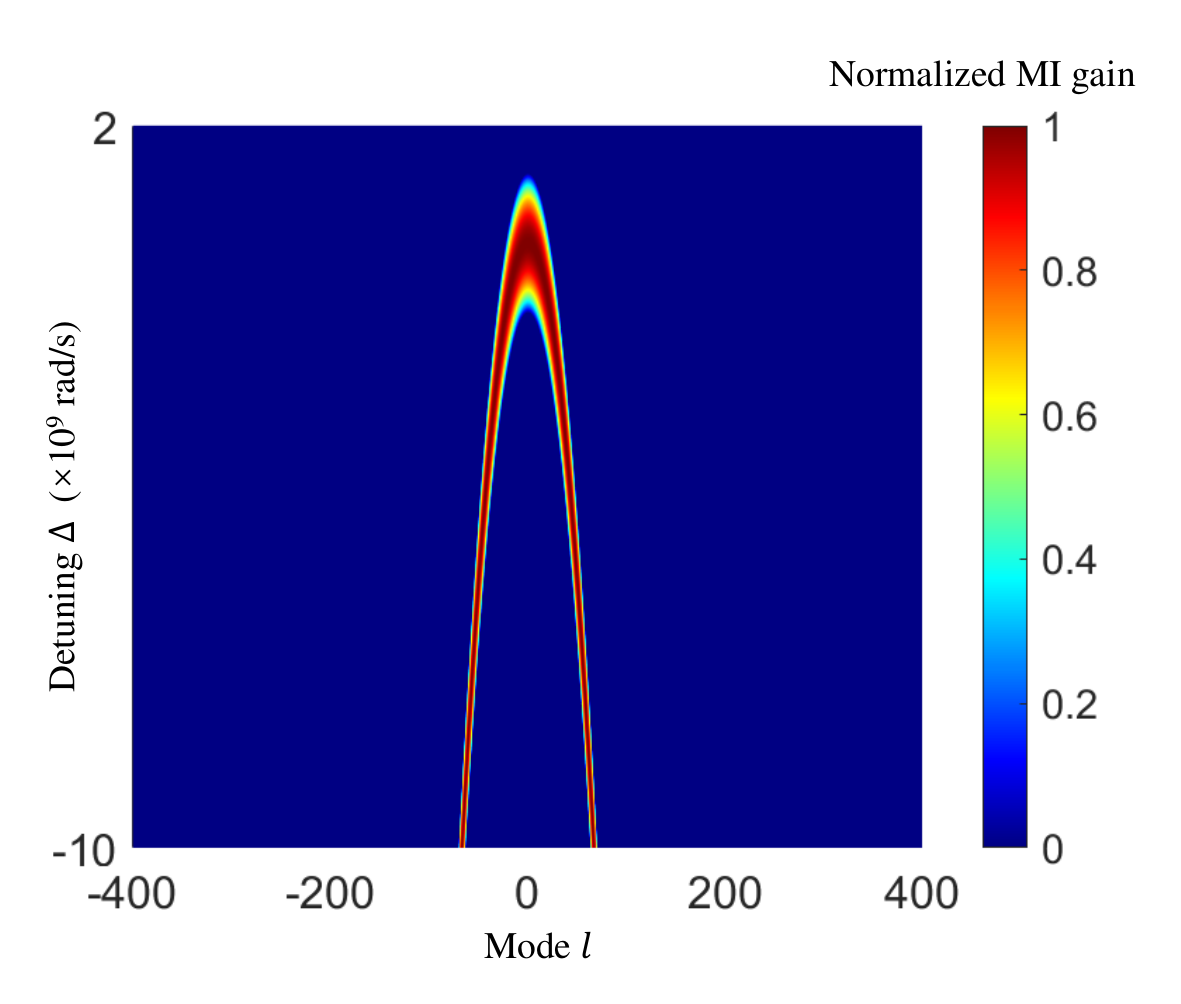}
    \caption{ MI gain of the trivial zero solution as a function of mode number and  detuning $\Delta$. }
\end{figure}
Similarly, the trivial zero solution demonstrates MI when the eigenvalues have a positive real part.
\begin{equation}
\lambda_{\pm} = -\Gamma \pm \sqrt{ \frac{2\gamma g_0^2}{\nu_f^2} B_{\mathrm{in}}^2 - \left( \Delta - \frac{k''_1L\nu_f}{2} \Omega^2 \right)^2 }.
\end{equation}
The instability of the trivial solution is evidently not governed by the detuning parameter but is instead induced by group velocity dispersion. Figure B.1 illustrates the MI gain associated with the trivial solution.

\section{Analytical Bloch-Messiah Decomposition}
\setcounter{figure}{0}
By integrating the methodologies proposed in Ref.~\cite{PhysRevLett.125.103601} and Ref.~\cite{HOUDE2024497}, the Analytical Bloch-Messiah Decomposition (ABMD) of  \( S_x(\omega) \) (the propagation matrix after similarity transformation) can be carried out through a systematic analytical procedure centered at \( \omega=0 \), and extended across the frequency domain. The process begins with the polar decomposition of the zero-frequency transformation
\begin{equation}
S_x(0) = P Y,
\end{equation}
where \( P \) is a real, symmetric, positive-definite symplectic matrix, and \( Y \) is a real orthogonal matrix. This decomposition serves as the foundation for identifying the canonical modes.

The matrix \( P \) is then partitioned into block form,
\begin{equation}
P = \begin{pmatrix}
A & B \\
B^T & C
\end{pmatrix},
\end{equation}
which allows the construction of a complex symmetric matrix
\begin{equation}
M = \frac{1}{2} \left( A - C + \mathrm{i}(B + B^T) \right).
\end{equation}
The Takagi (or Autonne) decomposition is then performed on \( M \), yielding
\begin{equation}
M = W \Lambda W^T,
\end{equation}
where \( \Lambda \) is diagonal with non-negative entries and \( W \) is a unitary matrix. This decomposition can be explicitly computed via the singular value decomposition
\begin{equation}
M = O \Lambda Q^\dagger,
\end{equation}
followed by constructing the unitary matrix
\begin{equation}
W = O \cdot \sqrt{ (O^T Q)^* }.
\end{equation}

With \( W \) obtained, we build the orthogonal symplectic matrix \( U \) as
\begin{equation}
U = \begin{pmatrix}
\operatorname{Re}(W) & -\operatorname{Im}(W) \\
\operatorname{Im}(W) & \operatorname{Re}(W)
\end{pmatrix},
\end{equation}
and defines the diagonal matrix \( \Xi = \Lambda + \sqrt{\Lambda\Lambda+E} \), from which the full diagonal scaling matrix
$
D = \begin{pmatrix}
\Xi & 0 \\
0 & \Xi^{-1}
\end{pmatrix}
$is constructed. The final component of the canonical form is given by
\begin{equation}
V^\dagger = U^T Y,
\end{equation}
completing the decomposition
$
S_x(0) = U(0) D(0) V^\dagger(0)
$
at zero frequency.

To extend this decomposition perturbatively around \( \omega = 0 \), the derivative of the symplectic transformation is computed as
\begin{equation}
S_x'(0) = \lim_{\delta \omega \to 0} \frac{S_x(\delta \omega) - S_x(0)}{\delta \omega},
\end{equation}
and used to define the auxiliary matrix
\begin{equation}
Q(0) = U^\dagger(0) S_x'(0) V(0).
\end{equation}
This matrix is then partitioned into four submatrices:
\begin{equation}
Q(0) = \begin{pmatrix}
Q_1 & Q_3 \\
Q_2 & Q_4
\end{pmatrix}.
\end{equation}
From these submatrices, the Hermitian and anti-Hermitian generator matrices \( H \) and \( K \) are constructed. The elements of \( H_1 \) and \( K_1 \) satisfies
\begin{equation}
\begin{cases}
(H_1)_{ij} = \dfrac{(Q_1)_{ij}(D)_{jj} + (Q_1)^{*}_{ji}(D)_{ii}}{(D)_{jj}^2 - (D)_{ii}^2}, & (i \ne j), \\[10pt]
(K_1)_{ij} = \dfrac{(Q_1)_{ij}(D)_{ii} + (Q_1)^{*}_{ji}(D)_{jj}}{(D)_{jj}^2 - (D)_{ii}^2}, & (i \ne j), \\[10pt]
(H_1)_{ii} - (K_1)_{ii} = \dfrac{(Q_1)_{ii} - (Q_1)^{*}_{ii}}{2(D)_{ii}}, & \\[10pt]
(K_1)_{ii} = 0. &
\end{cases}
\end{equation}

To compute \( H_2 \) and \( K_2 \), we solve the coupled equations:
\begin{equation}
\begin{cases}
Q_2 &= H_2 \Xi^{-1} - \Xi K_2, \\
Q_3 &= -H_2 \Xi + \Xi^{-1} K_2,
\end{cases}
\end{equation}
\begin{figure*}[htb]
    \centering
    \includegraphics[width=1\textwidth]{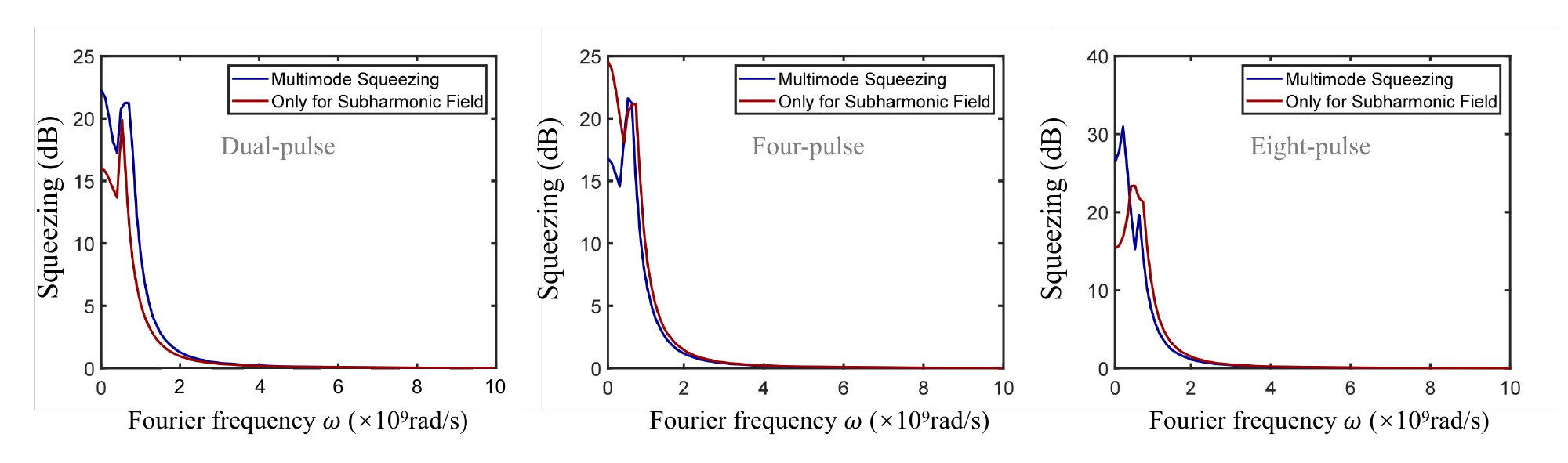}
    \caption{ Relationship between multimode squeezing and Fourier frequency for 2, 4, and 8 pulses within a single cavity round-trip period. }
\end{figure*}
which yield the remaining components of the generator matrices. Combining these, we construct
\begin{equation}
H = \begin{pmatrix}
H_1 & H_2 \\
-H_2 & H_1
\end{pmatrix}, \quad
K = \begin{pmatrix}
K_1 & K_2 \\
-K_2 & K_1
\end{pmatrix}.
\end{equation}
The variation of the diagonal matrix \( D \) with frequency is determined from the diagonal elements of \( Q_1 \), using the expression
\begin{equation}
(D)'_{ii}(0) = \frac{1}{2} \left[ (Q_1)_{ii} + (Q_1)_{ii}^* \right],
\end{equation}
which enables extrapolation of \( D(\omega) \) in the vicinity of \( \omega = 0 \).

To update the transformation matrices for nearby frequencies, a perturbative Magnus expansion is applied. The matrices \( U(\omega + \delta \omega) \) and \( V(\omega + \delta \omega) \) are approximated as
\begin{equation}
\begin{cases}
U(\omega + \delta \omega) &\approx U(\omega) e^{H(\omega) \delta \omega}, \\
V(\omega + \delta \omega) &\approx V(\omega) e^{K(\omega) \delta \omega},
\end{cases}
\end{equation}
which allows the decomposition
$
S_x(\omega) = U(\omega) D(\omega) V^\dagger(\omega)
$
to be analytically propagated across the frequency domain through iterative application of the differential steps.

According to the ABMD, optimal squeezed states can be achieved by appropriately tailoring either the output or input cavity modes. Figure C.1 presents a spectral comparison between the multimode squeezing levels obtained under different conditions and those derived by considering only the subharmonic field. Remarkably, the multimode squeezing achieved by considering solely the subharmonic modes can be substantial—occasionally even surpassing the squeezing obtained when all modes are included. This counterintuitive result arises because incorporating additional, non-squeezed modes may introduce greater squeezing degradation.

%%       \citet{<label>} ==> Lamport [21]
%%       \citep{<label>} ==> [21]
%%



\bibliography{ref.bib}

\end{document}